\documentclass[aps,twocolumn,showpacs,preprintnumbers,nofootinbib,prd,superscriptaddress,groupedaddress,10pt]{revtex4-1}

\makeatletter
\def\l@subsubsection#1#2{}
\def\l@subsubsubsection#1#2{}
\makeatother

\usepackage{footmisc}
\usepackage{amsmath}
\usepackage{graphicx,amssymb,amsmath,amsthm,amsfonts,epsfig,epsf}
\usepackage[usenames,dvipsnames,svgnames,table]{xcolor}
\usepackage{epstopdf}
\definecolor{darkred}{rgb}{0.5,0,0}
\usepackage{aas_macros}
\usepackage{amssymb}
\usepackage{array}
\usepackage{bm}
\usepackage{dcolumn}
\usepackage[utf8]{inputenc}
\usepackage{latexsym}
\usepackage{rotating}
\usepackage{longtable}

\usepackage{enumerate}
\usepackage{tensor,multirow}
\usepackage{braket}
\usepackage{url}
\usepackage[linktocpage,hidelinks]{hyperref}
\usepackage{footmisc}

\def\be{\begin{equation}}
\def\ee{\end{equation}}
\newcommand{\beq}{\begin{eqnarray}}
\newcommand{\eeq}{\end{eqnarray}}

\def\ba{\begin{align}}
\def\ea{\end{align}}

\newcommand{\F}{\mathcal{F}}

\DeclareMathOperator\erf{erf}

\begin{document}

\title{A covariant formulation of Relativistic Mechanics}

\author{Miguel Correia }
\email{mcorreia.alfra@gmail.com}

\affiliation{CERN, Theoretical Physics Department, 1211 Geneva 23, Switzerland and\\
Fields and Strings Laboratory, Institute of Physics, École polytechnique fédérale de Lausanne, Switzerland} 

\begin{abstract}
Accretion disks surrounding compact objects, and other environmental factors, deviate satellites from geodetic motion. Unfortunately, setting up the equations of motion for such relativistic trajectories is not as simple as in Newtonian mechanics. The principle of general (or Lorentz) covariance and the mass-shell constraint make it difficult to parametrize physically adequate 4-forces. Here, we propose a solution to this old problem. We apply our framework to several conservative and dissipative forces. In particular, we propose covariant formulations for Hooke's law and the constant force, and compute the drag due to gravitational and hard-sphere collisions in dust, gas and radiation media. We recover and covariantly extend known forces such as  Epstein drag, Chandrasekhar's dynamical friction and Poynting-Robertson drag. Variable-mass effects are also considered, namely Hoyle-Lyttleton accretion and the variable-mass rocket. We conclude with two applications: 1. The free-falling spring. We find that Hooke's law corrects the deviation equation by an effective Anti-de Sitter tidal force; 2. Black hole infall with drag. We numerically compute some trajectories on a Schwarzschild background supporting a dust-like accretion disk.
\end{abstract}

\maketitle

\tableofcontents

\section{Introduction}\label{Introduction}
\label{sec:intro}

The observation of gravitational waves brought unprecedented experimental access to ultra-relativistic macroscopic systems: binaries of black holes (BHs) and other compact objects \cite{LIGOScientific:2016aoc,Barack:2018yly}. In the coming years, LISA \cite{LISA,Ballmer:2022uxx} is projected to access a lower frequency range enough to start observing extreme-mass-ratio inspirals (EMRIs). An EMRI consists of a system of two inspiralling BHs with a large mass discrepancy. In an EMRI the smaller BH can be seen as a point-particle which practically follows a geodesic on the metric that would be purely generated by the larger BH. There is a slight deviation from the geodesic due to gravitational wave emission and associated back-reaction \cite{Barack:2018yvs,Poisson:2011nh}. 
\par
Candidates for EMRIs consist of inspiralling stellar mass compact objects into super massive BHs which are believed to reside in galactic centers  and active galactic nuclei  \cite{kormendy}. It is likely that such EMRIs do not evolve in a true vacuum due to the presence of matter, namely accretion disks of gas or dust, or dark matter halos \cite{Levin:2003kp,frank,1995PASP,Donoso:2013qsu}. In such environments the infalling stellar mass BH gravitationally disturbs surrounding matter leading to an overall drag force on the BH itself (an effect typically known as \emph{dynamical friction} \cite{Chandra,Rephaeli,Petrich,Ostriker:1998fa,Barausse:2007ph}). Closer encounters lead instead to accretion of matter, wherein the BH increases its mass via gravitational capture \cite{1939PCPS...35..405H,1944MNRAS.104..273B,1952MNRAS.112..195B,Petrich}. Such environmental effects may leave an imprint on the gravitational wave signal of certain EMRIs \cite{Barausse:2014pra}.
\par
Naturally, objects without an event horizon, which may range in mass from neutron stars and exotic compact objects \cite{Barack:2018yly} to asteroids and dust grains, may also interact with matter via direct contact collisions, as in typical atmospheric or hydrodynamic drag \cite{PhysRev.23.710,Weidenschilling1977AerodynamicsOS,1979ApJ...231...77D,Hoang:2017ijx}. 
\par
Environmental effects are not exclusive to massive media. It is well known that radiation exerts an outward pressure on the orbital motion of dust grains and other small particles around stars \cite{1979Icar...40....1B,1995Icar..116..186L,2013MNRAS.436.2785K}. Radiation also has a dissipative impact on the orbits of such objects, an effect known as Poynting-Robertson drag \cite{1904RSPTA.202..525P,1937MNRAS..97..423R,1950ApJ...111..134W,Rafikov:2011uv}. It is equally possible that in systems within or close to Eddington luminosity, such as quasars or active galactic nuclei, the orbits of compact objects may be disturbed by their own gravitational influence on the radiation field (an effect akin to dynamical friction) \cite{Syer:1994vr}.
\par
These are astrophysical scenarios where the geodesic equation needs correction because satellite motion is being perturbed or driven away by external forces. While currently more of an academic interest, one may also consider other mechanical problems in a curved background, where the same holds true. One such example, which we will address here, consists of attaching an elastic spring to a pair of masses and let the system fall gravitationally, e.g. into a BH. Given an external, non-gravitational, effect on the motion of a test-particle, we will be interested on finding the appropriate equations of motion. As we will try to convey in what follows, this procedure is not as simple as in Newtonian mechanics.
\par
If $m$ and $\tau$ are, respectively,  the mass and proper time of a test-particle  that is moving with 4-velocity $u^\mu$ on a curved background with metric $g_{\mu \nu}$,  one typically writes Newton's second law in curved space-time as
\par
\be
\label{eq:NLo}
m {D u^\mu \over d \tau} = m\left({d u^\mu \over d \tau} + \Gamma^{\mu}_{\;\nu\sigma}u^{\nu}u^{\sigma}\right) = f^\mu,
\ee
where $f^\mu$ is the 4-force acting on the test-particle and $\Gamma^{\mu}_{\;\nu\sigma}$ is the connection associated with the metric $g_{\mu \nu}$. Naturally, if there is no force acting on the test-particle we have $f^\mu = 0$ and eq. \eqref{eq:NLo} reduces to the geodesic equation. Our task is then to find which $f^\mu$ better parametrizes a given external effect.
\par
Importantly, the 4-velocity $u^\mu$ is constrained to be a unit timelike vector, 
\be
\label{eq:ms}
u^2 = g_{\mu \nu} u^\mu u^\nu = - 1,
\ee
where we took the `mostly plus' signature for the metric. This condition is sometimes called the \emph{mass-shell} constraint, which the solution of \eqref{eq:NLo} must satisfy. Differentiating eq. \eqref{eq:ms} and using \eqref{eq:NLo} leads to the following constraint on the 4-force $f^\mu$,
\be
\label{eq:fp0}
f \cdot u = g_{\mu \nu} f^\mu u^\nu = 0.
\ee
This means that any physically adequate choice of $f^\mu$, i.e. which conserves the mass-shell \eqref{eq:ms}, should be orthogonal to the momentum $u^\mu$ at all times. 
\par
In Newtonian mechanics, the force is an arbitrary object, which can be parametrized according to a given external effect on the test-particle's motion. Unfortunately, such intuitive reasoning does not typically carry onto the 4-force $f^\mu$. Most obvious generalizations of the Newtonian force to the  4-force $f^\mu$ do not satisfy the orthogonality constraint \eqref{eq:fp0}. For example, the naive covariant extension of Hooke's law, $f^\mu= -k \,\Delta x^\mu$, where $k$ is the spring constant and $\Delta x^\mu$ the 4-displacement of the spring, would not obey \eqref{eq:fp0} at all times.\footnote{Such a choice would lead to the  unphysical (in our view) dependence of the rest-mass $m$ on the spring constant $k$ (see e.g. \cite{PhysRevD.6.1474}).}  Another example would be a generic drag 4-force $f^\mu = - b \, u^\mu$, where $b$ is some drag coefficient. Now we have $f \cdot u = b$ which can only satisfy \eqref{eq:fp0} in the trivial case $b = 0$.
\par
Of course, in flat space-time, one may just forget about covariance and work directly with the force $\vec{F}$ in the special relativistic generalization of Newton's second law,
\be
\label{eq:NLn}
m {d (\gamma \vec{v} )\over d t} = \vec{F} \qquad \text{ with } \qquad \gamma = {1 \over \sqrt{1 - v^2/c^2}},
\ee
where $\vec{v}$ the velocity of the test-particle and $c$ the speed of light. In this equation, $\vec{F}$ should be the same physical object as in the original Newton's second law and can therefore be parametrized in the same way. The only difference with respect to the Newtonian case is the presence of the $\gamma$ factor which increases the inertia of the test-particle when $v \to c$ and prevents it from surpassing the speed of light $c$
\par
Well studied examples with exact solutions include the constant force case, $\vec{F} = $ constant, which leads to \emph{hyperbolic motion} \cite{1909AnP...335....1B,1910AnP...338..649S,Misner:1973prb}, and the \emph{relativistic harmonic oscillator} \cite{1957AmJPh..25..535M,1965Natur.205..892H,1994AmJPh..62..531M} which follows from Hooke's law $\vec{F} = - k \, \Delta \vec{x}$. Both examples obviously reduce to the corresponding Newtonian cases at non-relativistic velocities. In the constant force case, relativistic effects necessarily appear at late times, for which $v \to c$ (instead of $v \to \infty$ in the Newtonian case). In the case of Hooke's law, trajectories with large amplitudes, which would lead to faster-than-light motion in Newtonian mechanics at the equilibrium point, are instead ``flattened" near this point, where relativistic velocities achieved instead. In the ultra-relativistic limit, the wordline assumes a zig-zag or `saw-like' shape,\footnote{This was recently observed in an optical lattice that simulates a relativistic harmonic oscillator, namely on an energy band with the mass-shell \eqref{eq:ms} energy-momentum dispersion but a much smaller ``speed of light" $c = 143$ mm/s \cite{Fujiwara:2017lkd}.} akin to a photon bouncing between two parallel mirrors. See e.g. \cite{Goldstein} for a pedagogical introduction to both cases.
\par
The 4-force $f^\mu$ can be found in terms of $\vec{F}$, as explained in many textbooks \cite{synge1965relativity,Mller1972-MLLTTO,Landau:1975pou,rindler2001relativity,rahaman2014special}. The idea is to match the spatial part of eq. \eqref{eq:NLo}, in flat space-time, with eq. \eqref{eq:NLn}, making use of the usual relation between coordinate and proper times $d t = \gamma d \tau$ and $u^\mu = \gamma(1, \vec{v})$ (we set $c=1$ from now on). This fixes the spatial part of the 4-force $\vec{f}$, and the time component $f^0$ follows from the orthogonality condition \eqref{eq:fp0}. The result is
\be
\label{eq:fF}
f^\mu = \gamma \left(\vec{F} \cdot \vec{v} ,  \vec{F} \right).
\ee
\par
A similar strategy can of course be applied in curved-space time, where one directly parametrizes $\vec{f}$ in a given curved background and fixes $f^0$ using \eqref{eq:fp0}. However, by treating $\vec{f}$ and $f^0$ separately one is not obeying the principle of covariance and, besides going against the underlying spirit of the theory of relativity, there is no guarantee that the resulting expression for $f^\mu$ transforms covariantly, i.e. as a 4-vector. Moreover, it is $\vec{F}$ that is the force, not $\vec{f}$, so one must be careful in choosing $\vec{f}$ as to correspond to the physically appropriate force $\vec{F}$ in flat space-time.
\par
In essence, parametrizing the 4-force in a covariant and physically meaningful way such that the mass-shell constraint \eqref{eq:ms} is also satisfied does not seem to be an obvious task. Of course, as a purely mathematical problem this should not be too hard, the point is if typical Newtonian forces can be described covariantly, i.e. whether the intuitive Newtonian picture that is part of every physicist's training translates in some way to $f^\mu$. Concretely, the question we are asking is: Given a Newtonian force, for instance Hooke's law or a drag force, is there a corresponding covariant 4-force $f^\mu$ that describes it and which is also orthogonal to $u^\mu$ at all times? 
\par
To our knowledge, despite the seniority of the subject, there is yet no covariant framework for relativistic forces that is capable of addressing this question. In fact, such absence may somewhat explain the relatively small amount of relativistic studies on the aforementioned astrophysical problems \cite{Barausse:2007dy,Gair:2010iv,Bini:2008vk,DeFalco:2018vgk}, as most work takes a pseudo-Newtonian (or even purely Newtonian) approach (see e.g. \cite{1993MNRAS.265..365V,Narayan:1999ay,Karas:2001vq,Kocsis:2011dr,Cardoso:2020iji,Peng:2021vzr} and references therein). Indeed, the vast majority of  environmental effects on the motion of astrophysical bodies  do not seem to have covariant descriptions. One notable exception and, to our understanding, the only known fully covariant example, is the Poynting-Robertson 4-force which, in its simplest form as derived by Robertson in 1937 \cite{1937MNRAS..97..423R}, reads
\be
\label{eq:RPintro}
f^\mu = - e \sigma (u \cdot n) \left[ (u \cdot n) u^\mu + n^\mu \right],
\ee
where $e$ is the radiation energy density, or the \emph{intensity} of radiation in $c=1$ units, on the rest-frame of a source (e.g. a star), $\sigma$ is the spherical cross-section of the test-particle and $n^\mu$  is a null vector which specifies the direction of the radiation flux.
\par
It is clear that eq. \eqref{eq:RPintro} satisfies the mathematical requirements of being both covariant and orthogonal to $u^\mu$. However, while expression \eqref{eq:RPintro} covariantly describes a force caused by absorption and scattering of radiation and, in particular, quantitatively explains the Poynting-Robertson effect, there is nothing immediately intuitive about its form \eqref{eq:RPintro}. It is not surprising that the ``nature [of the Poynting-Robertson force] has been the subject of considerable
controversy and misunderstanding since
the beginning of the [twentieth] century" \cite{1979Icar...40....1B}. While the quoted work from 1979 by Burns, Lamy and Soter seems to have settled much of the discussion on the physical origin of the Poynting-Robertson effect, their exposition still faced some criticism in recent years \cite{2014Icar..232..249K}.\footnote{See \cite{2014Icar..232..263B} for a response from the original authors.} It is also unfortunate that the Poynting-Robertson 4-force has been unjustifiably (and, as we will see, erroneously) used to covariantly describe drag due to collisions with dust/gas particles, where $e$ and $n^\mu$ in eq. \eqref{eq:RPintro} were interpreted, respectively, as the proper mass density and 4-velocity of the dust/gas medium \cite{Bini:2013pui,Bini:2013es,Bini:2016tqz}. Such confusion, in our view, although in part caused by the absence of a direct Newtonian analogue for the Poynting-Robertson effect, is another symptom of the absence of a covariant framework for relativistic forces.
\par
Historically, as far as we understand, much of the endeavor on the search for such a formulation has been specific to the constant force. Naturally, a constant $f^\mu$ cannot satisfy the orthogonality constraint \eqref{eq:fp0} at all times $\tau$. Instead, as noted early by Born \cite{1909AnP...335....1B}, in 1909, hyperbolic motion has constant \emph{proper} acceleration. That is, if $\vec{F}$ is constant and aligned with $\vec{v}$ in eq. \eqref{eq:fF} it follows that $f^\mu$, despite being variable, has a constant norm, $f^2 = |\vec{F}|^2 = \text{constant}$. Then, eq. \eqref{eq:NLo} implies that $(du^\mu / d\tau)^2 = \text{constant}$. The latter condition can be easily turned into the covariant statement $(D u^\mu / d \tau)^2 = \text{constant}$, which may now be used to covariantly \emph{define} motion under a constant force. However, just specifiying that the 4-acceleration has a constant norm is not sufficient to describe the motion of a test-particle on a 4-dimensional manifold. Rindler \cite{rindler1960hyperbolic}, in 1960, proposed that this condition be supplemented with the requirement of \emph{planarity}, i.e. that the worldline be \emph{torsionless}. He showed that this would amount to two additional conditions that, together with constant norms for the 4-acceleration and the 4-velocity (i.e. the mass-shell \eqref{eq:ms}), would provide the necessary four equations that fix the evolution of a test-particle. One immediate question is, of course, how can the requirement of planarity be physically justified. It is also not clear how this approach generalizes to other forces, for example Hooke's law or drag forces. Different proposals that supplement, or arrive at, the constancy of the proper acceleration have been and continue to be suggested to this day \cite{marder1957uniform,gautreau1969constant,friedman2012making,friedman2013covariant,de2015uniformly,friedman2015uniform,olmo2018accelerated}. It is quite remarkable that the covariant formulation of the constant force, arguably the ``simplest" force, is still an active topic of research today.
\par
Instead of dealing directly with the 4-force or the equations of motion \eqref{eq:NLo}, one may see if better luck is found within a Lagrangian (or Hamiltonian) formulation. At first sight, an obvious advantage is that the Lagrangian is a scalar function that, under the requirement of covariance, should be the same in every frame, i.e. be a scalar invariant. The Lagrangian may then be built from contractions of $g_{\mu \nu}$, $u^\mu$ and other relevant covariant objects within the system in study. While this method works well for many topics, such as relativistic field theory \cite{Landau:1975pou,Weinberg:1995mt}, here we still have to impose the non-holonomic constraint of the mass-shell \eqref{eq:ms}. Motivated, in part, by the search for a relativistic theory of quantum mechanics, much thought was given in the 1950s and early 1960s on how to incorporate \eqref{eq:ms} into a variational method, namely by Dirac \cite{dirac1950generalized,dirac1958generalized} and contemporaries of his (see \cite{hannibal1991hamilton} and references therein). Unfortunately, significant difficulties are also encountered here. To quote a few authors: ``It seems to be established by now that relativistic dynamics is marred by the impossibility of translating it into terms
of a Hamiltonian formalism." \cite{kalman1961liouville} or ``[The equations of motion for the relativistic harmonic oscillator] may be derived from a variational principle, but not in an unambiguously Lorentz-invariant fashion." \cite{PhysRevD.6.1474}.
Finally, to quote \cite{schay1962relativistic}: "It is a strange fact that although the theory of relativity is almost sixty years old, no universally accepted covariant generalization of the Euler-Lagrange-Hamilton-Jacobi theory of mechanics has been developed." 
\par
It appears that another sixty years have passed and the situation has not improved dramatically. Though efforts in this direction seem to have waned after Currie, Jordan and Sudarshan's \emph{no-interaction theorem} \cite{RevModPhys.35.350} and consequent extensions \cite{currie1963interaction,doi:10.1063/1.1704121, leutwyler1965no}. As the name suggests, the no-interaction theorem states that under the requirement of Poincaré invariance a system of a finite number of particles must be free, i.e. every particle must follow a straight line.  Indeed, instantaneous action-at-a-distance is obviously forbidden due to finiteness of the speed of light, and if, instead, one considers retarded interactions, then there is a finite time interval in which energy-momentum conservation is violated. Unless, of course, a dynamical field carries and exchanges the energy-momentum variation in that time interval. In this case the no-interaction theorem does not apply given that a field has infinitely many degrees of freedom.
\par
This motivates the use of fields to mediate interactions between relativistic particles. However, only  vector fields, such as the electromagnetic field $A^\mu$, seem to preserve the mass-shell \eqref{eq:ms} \cite{kalman1961lagrangian, schay1962relativistic}. The corresponding Euler-Lagrange equations then fix the 4-force to be given by Lorentz' law
\be
\label{eq:Lorentz}
f^\mu = \F^{\mu \nu} u_\nu,
\ee
where $\F^{\mu \nu} = \partial^\mu A^\nu - \partial^\nu A^\mu$ is the Faraday tensor.
\par
In summary, the Lorentz 4-force \eqref{eq:Lorentz} looks like  the only choice for $f^\mu$ which is compatible with the principles of relativity. Indeed, apart from gravity, which is encoded in the metric $g_{\mu \nu}$, all other known fundamental forces of nature are vectorial. Therefore, there should be no need to search for other 4-forces as all classical natural phenomena should have a description in terms of eq. \eqref{eq:NLo}, Lorentz' law \eqref{eq:Lorentz} and associated field equations. While this is technically true, in practice it may not always be the case. Obviously, the Lorentz force \eqref{eq:Lorentz} has had myriads of applications across the last century and until the present day. However, as mentioned in the beginning, today there is also growing experimental evidence on \emph{macroscopic} relativistic systems which may not evolve in a true vacuum, and for which environmental effects may play a role in the dynamics. Describing such environmental effects in terms of eqs. \eqref{eq:NLo} and \eqref{eq:Lorentz} seems completely unfeasible in practice. Rather, a \emph{phenomenological} description, which is agnostic to the degrees of freedom of the environment, should be the most appropriate.
\par
In looking for an alternative, non-fundamental, description the no-interaction theorem should be completely mute. The mechanical system in study should simply consist of the test-particle, which is nonetheless being acted by an external agent. In particular, such a system will not conserve energy-momentum, a key ingredient of the no-interaction theorem.\footnote{Naturally, the full system of test-particle + environment should still conserve energy-momentum. } This idea is obviously central to Newton's original formulation of his second law, for which one may freely parametrize the force to model the action of an external agent whose fundamental origin may be completely unknown. It is entirely possible that we have missed some important references, given that the theory of relativity is more than a century old, but we are not aware of a corresponding formalism for relativistic forces that, again, is covariant and preserves the mass-shell \eqref{eq:ms}.
\par
In this work we propose such a framework. We now outline the rest of the paper and briefly summarize our method. We start, in section \ref{sec:eom}, by proving that any 4-force $f^\mu$ that preserves the mass-shell \eqref{eq:ms} can always be written in Lorentz' law form \eqref{eq:Lorentz} where $\F^{\mu \nu}$ is an arbitrary antisymmetric tensor.\footnote{It is trivial that antisymmetry of $\F^{\mu \nu}$ implies the orthogonality constraint \eqref{eq:fp0} and therefore conservation of \eqref{eq:ms}. Here we also prove the converse statement.} In section \ref{sec:covmap}, we connect $\F^{\mu \nu}$ with a Newtonian description. We write
\be
\label{eq:2Fout} 
\F^{\mu \nu} = U^\mu F^\nu - U^\nu F^\mu,
\ee
where $U^\mu$ is a unit timelike 4-vector and $F^\mu$ is an arbitrary 4-vector. We show through the use of the equivalence principle that $F^\mu$ maps directly to the force in Newton's second law \eqref{eq:NLn}. Concretely, if $U^\mu$ is associated with the 4-velocity of some object, then $\vec{F}$ is the force on the instantaneous local Lorentz rest-frame of that object.\footnote{I.e. $f^\mu$ given by eqs. \eqref{eq:Lorentz} and \eqref{eq:2Fout} reduces to eq. \eqref{eq:fF} in the local Minkowski frame in which $U^\mu = (1, \vec{0})$. The existence of such a frame is guaranteed by the equivalence principle.}
 In this way, we find the covariant generalization of several conservative (section \ref{sec:lcv}) and dissipative forces (section \ref{sec:cf}).
\par
For example, identifying $U^\mu$ with the 4-velocity of a point charge in uniform motion, implies that $\vec{F}$ is the electric force on the rest-frame of the charge, i.e. Coulomb's law. The associated covariant Faraday tensor \cite{jackson_classical_1999} then follows from eq. \eqref{eq:2Fout}. Other examples include Hooke's law and drag forces,
\be
\label{eq:HD}
F_\text{Hooke}^\mu = - k \,\Delta x^\mu, \qquad F_\text{Drag}^\mu = - B u^\mu,
\ee
where $B= B(u^\mu,  U^\mu)$ is some model-dependent drag coefficient. In the case of drag, $U^\mu$ is associated with the 4-velocity of the medium, while for Hooke's law, $U^\mu$ can be taken as the 4-velocity of whatever object is attached to the other end of the spring.
\par
The equivalence principle relates $B$ with typical drag coefficients in Newtonian mechanics. However, since most of these drag coefficients were originally derived within a Newtonian setting, e.g. Newtonian hydrodynamics, one should first re-derive these drag coefficients within a special relativistic setting before covariantly generalizing them.\footnote{Or, in alternative, understand their Newtonian regime of validity in a covariant way.} For this reason, section \ref{sec:cf} is supplemented with a special relativistic derivation of a class of dissipative forces. Namely, those for which free molecular flow applies (i.e. with large Knusden number). In subsection \ref{ssec:scat}, this is done for several of the drag forces already mentioned and the associated covariant drag coefficients $B$ are found (see table \ref{table:B}). In subsection \ref{sec:vms}, we extend our formalism to variable mass systems. In particular, Hoyle-Lyttleton accretion and the variable-mass-rocket are given covariant descriptions.
\par
We consider a couple of simple applications in sec. \ref{sec:app}. We make  use of eq. \eqref{eq:HD} to respectively study an elastic spring in free-fall and the infall of an observer onto a black hole with an accretion disk.  See figure \ref{fig:orbits} for some explicit non-geodetic trajectories on a Schwarzschild background.
\par
In section \ref{sec:conclusion} we summarize our findings, the limitations of our method and discuss some future directions.
\par

\section{Equations of motion}
\label{sec:eom}

We consider units where $m=c=1$. We take space-time as the four dimensional pseudo-Riemannnian manifold endowed with a metric $g_{\mu \nu}$. We take the `mostly plus' convention for which the Minkowski metric reads
\be
\eta_{\mu \nu} = \mathrm{diag}(-1,+1,+1,+1).
\ee
We make use of the usual notations for the temporal and spatial parts of a 4-vector, $A^\mu = (A^0, \vec{A})$ with $\vec{A} = (A^1, A^2 , A^3)$, the relation between covariant and contravariant tensors,  $A_\mu \equiv g_{\mu \nu} A^\nu$ and the dot product, $A \cdot B \equiv A_\mu B^\mu$ with the special case $A^2 \equiv A_\mu A^\mu$. 
\par
The line element of the metric reads
\be
- d \tau^2 = g_{\mu \nu} dx^\mu dx^\nu,
\ee
for which a massive particle respects $d \tau^2 > 0$. Then, the 4-velocity $u^\mu \equiv {dx^\mu \over d \tau}$ is restricted to the mass-shell
\be
\label{eq:mass-shell}
u^2 = g_{\mu \nu} u^\mu u^\nu = -1.
\ee
This condition ensures that only $3$ of the $4$-velocity components are independent. Differentiating the above leads to
\be
\label{eq:mass-shell2}
g_{\mu \nu} \dot{u}^\mu  u^\nu + {1 \over 2} u^\sigma u^\mu u^\nu \partial_\sigma g_{\mu \nu} = 0,
\ee
which in terms of the $4-$acceleration,
\be
\label{eq:eom}
a^\mu \equiv {D u^\mu \over d \tau} = \dot{u}^\mu + \Gamma^\mu_{\; \rho \sigma}  u^\rho u^\sigma,
\ee
where
\be
 \Gamma^\mu_{\; \rho \sigma} \equiv {1 \over 2} g^{\mu \nu} ( \partial_\rho g_{\nu \sigma} +  \partial_\sigma g_{\nu \rho}  - \partial_\nu g_{\rho \sigma} ),
\ee
are the Christoffel symbols, reads
\par
\be
\label{eq:fu}
a_\mu u^\mu = 0.
\ee
Now, given an arbitrary 4-vector $\bar{a}^\mu$ we may project out the component along $u^\mu$ to construct an orthogonal vector to $u^\mu$.
Thus, $a^\mu$ takes the general form
\be
\label{eq:projection}
a^\mu = \bar{a}^\mu + (\bar{a}^\nu u_\nu)  u^\mu,
\ee
for any $4$-vector $\bar{a}^\mu$. Making further use of \eqref{eq:mass-shell} we may also write \eqref{eq:projection} as
\be
\label{eq:F}
a^\mu =\F^{\mu \nu} u_\nu
\ee
with
\be
\label{eq:F2}
\F^{\mu \nu} \equiv u^\mu \bar{a}^\nu - u^\nu \bar{a}^\mu + \; \epsilon^{\mu \nu \rho \sigma} \omega_\rho u_\sigma,
\ee
for any $\omega^\mu$. Note that the $\omega^\mu$ dependent piece is orthogonal to $u^\nu$ so it drops out of \eqref{eq:F}. Importantly, any antisymmetric tensor can be written in the form \eqref{eq:F2}. Each of $\bar{a}^\mu$ and $\omega^\mu$ contribute with 3 independent components (the component along $u^\mu$ cancels out of \eqref{eq:F2}) that make up the 6 independent components of a generic antisymmetric tensor.\footnote{For concreteness, take a local inertial frame where $u^\mu = (1,\vec{0})$. Then, the spatial components of $\bar{a}^\mu$ and $\omega^\mu$, can be respectively seen as the ``electric" and ``magnetic" fields of $\F^{\mu \nu}$. The time components are proportional to $u^\mu$ and drop out of \eqref{eq:F2}.} Hence, eq. \eqref{eq:fu} implies
\be
\boxed{
\label{eq:law}
\dot{u}^\mu + \Gamma^\mu_{\; \rho \sigma}  u^\rho u^\sigma = \F^{\mu \nu} u_\nu}
\ee
for any $\F^{\mu \nu} = - \F^{\nu \mu}$.
\par 
Conversely, starting from \eqref{eq:law} with an antisymmetric $\F^{\mu \nu}$ and contracting with $u_{\mu}$ makes the RHS vanish, which implies eq. \eqref{eq:fu} and thus $u^2 = \text{constant}$. The constant should be fixed to $1$ as an initial condition to \eqref{eq:law}. This establishes equivalence of eq. \eqref{eq:mass-shell} with eq. \eqref{eq:law} and the initial condition $u^2(0) = - 1$.
\par
To a given choice of $\F^{\mu \nu}$ corresponds a certain parametrization of a non-gravitational force. Note that $\F^{\mu \nu}$ need not be a field, it may depend on the velocity. Naturally, when $\F^{\mu \nu}$ is proportional to the electromagnetic field tensor, we recover the covariant form of Lorentz' force law, and in the absence of force, $\F^{\mu \nu} = 0$, eq. \eqref{eq:law} becomes the geodesic equation.

\section{Covariant map }
\label{sec:covmap}

To make practical use of \eqref{eq:law} we look for a map that given a Newtonian force $\vec{F}_N$ produces the corresponding $\F^{\mu \nu}$. This map should be such that \eqref{eq:law} reduces to Newton's second law,
\be
\label{eq:Fma}
{d \vec{u} \over dt} = \vec{F}_N, \qquad \text{with} \qquad \vec{u} = {\vec{v} \over \sqrt{1 - v^2}},
\ee
in a suitable local inertial frame, where $\vec{v}$ is the test-particle's velocity. Noting that $dt = u^0 d\tau$, we can rewrite the above as
\be
\label{eq:N2}
{d \vec{u} \over d \tau} = \vec{F}_N  u^0.
\ee
Thus, one option for $\F^{\mu \nu}$ is
\be
\label{eq:2F}
\F^{\mu \nu} \equiv U^\mu F^\nu - U^\nu F^\mu,
\ee
where $U^\mu$ is a unit timelike vector 
\be
\label{eq:msU}
U^2 = -1,
\ee
and $F^\mu$ is a 4-vector whose spatial part matches $\vec{F}_N$ in a local inertial frame where $U^\mu =(1,\vec{0})$. That is, 
\be
\label{eq:presc}
\left[ \, \vec{F} = \vec{F}_N \;\, \right]_{\vec{U} = 0}, 
\ee
in the instantaneous local inertial frame
\be
\label{eq:normal}
g_{\mu \nu}\left(x^\alpha(\tau)\right) = \eta_{\mu \nu}, \qquad \Gamma^\mu_{\; \rho \sigma}\left(x^\alpha(\tau) \right) = 0.
\ee
This is achieved, for example, in a Riemann normal coordinate system centered at the test particle's instantaneous position \cite{Misner:1973prb}. Once in Minkowski we are free to apply a Lorentz boost and align $U^\mu$ along the time direction. Hence, one can always find a frame where \eqref{eq:presc} and \eqref{eq:normal} are both applicable. This is consistent with the local flatness theorem \cite{Schutz:1985jx, Misner:1973prb} but, more generally, with the equivalence principle.
\par
It is now straightforward to check that in the frame \eqref{eq:normal}, the spatial part of \eqref{eq:law} with $\F^{\mu \nu}$ given by eq. \eqref{eq:2F} reduces to eq. \eqref{eq:N2}. Therefore, one can use \eqref{eq:presc} to find $F^\mu$ and plug into \eqref{eq:2F} to get $\F^{\mu \nu}$.\footnote{Specifying $\vec{F}$ in  the local frame \eqref{eq:normal} is enough to fix $\F^{\mu \nu}$ given that the component of $F^\mu$ along $U^\mu$ does not contribute to eq. \eqref{eq:2F}. Just like the Newtonian force, only $3$ components of $F^\mu$ independently contribute to the 4-force. \label{fn}}  Naturally, this is only useful if the expression for $\vec{f}$ is known in the frame \eqref{eq:normal} where $\vec{U} = 0$. Given that $U^2 = -1$ we may identify $U^\mu$ with the 4-velocity of some physical massive object. 
\par
One obvious option is to take this object as the test-particle itself, so that $U^\mu = u^\mu$, and $\vec{F}$ is the force on the test-particle in its own instantaneous inertial rest frame. There are some cases where this identification can be useful, such as with  dissipative forces (see sec. \ref{sec:cf}). Another example is the Abraham-Lorentz self-force \cite{jackson_classical_1999} which is, in fact, only strictly valid in the instantaneous rest frame of the accelerated charge. Application of eqs. \eqref{eq:2F} to \eqref{eq:normal} leads to Dirac's covariant expression for the corresponding 4-force \cite{1938RSPSA.167..148D} (see appendix \ref{app:ald}).
\par
Alternatively, one may consider $U^\mu$ as the 4-velocity of a secondary object. Note that, strictly speaking, $U^\mu$ is a vector on the tangent space at the test-particle's instantaneous position. This means that such a secondary object would have to be co-moving with the test-particle. Most drag forces, which are due to relative motion with respect to a fluid, fit into this description. In this case, $U^\mu$ can be interpreted as the 4-velocity of the fluid at the test-particle's instantaneous position and $\vec{F}$ is the force on the test-particle in the fluid's rest frame. In fact, for most forces, there is typically a secondary object which is responsible in some way by the force $\vec{F}$ acting on the test-particle. Say an electric charge that acts on the test-particle through a Coulomb field, the opposite end of an elastic spring to where the test-particle is attached or, as mentioned above, the element of fluid surrounding the test-particle at each moment. It is helpful to refer to such secondary object as the \emph{force applier}.
\par
If the force applier is not exactly co-moving with the test-particle but rather keeps itself in the test-particle's close victinity then the tangent spaces of both objects will approximately overlap and in this limit $U^\mu$ should retain its intepretation as the force applier's 4-velocity. In this case both test-particle and applier will share the same local inertial frame \eqref{eq:normal} where space is approximately flat and the equivalence principle holds.
\par
In flat space, there is a single tangent space that is common to every point, i.e. Minkowski space-time itself. In this case $U^\mu$ can be literally taken as the force applier's 4-velocity, even if test-particle and force applier are a finite distance apart. Moreover, most Newtonian forces between two separated objects depend on the Euclidean distance between them. In flat-space these ``at-a-distance" Newtonian expressions can be easily covariantly generalized using the Minkowski metric.  This is due to the fact that in flat space coordinates can be interpreted as 4-vectors.
The resulting expression for $\F^{\mu \nu}$ will not be \emph{general} covariant but rather \emph{Lorentz} covariant.  We apply this method in section \ref{sec:lcv} to find Lorentz covariant generalizations of typical Newtonian conservative forces. Namely, Coulomb's law, Hooke's law and the constant force.
\par
As just mentioned, Lorentz covariant expressions can be useful if test-particle and force applier are sufficiently close so that space in their victinity is approximately flat. In particular, if the applier is in free-fall, in the absence of force, the test-particle would also be in free-fall. Their separation would then be dictated by the geodesic deviation equation \cite{Misner:1973prb}. If there is force, however, the same cannot hold true since the test-particle will not be in free-fall. In this case the equation for the deviation between test-particle and force applier's worldlines needs correction. In section \ref{sec:ffs}, we find this correction explicitly for the case of an elastic spring connecting both objects using the Lorentz covariant generalization of Hooke's law.
\par
\section{Conservative forces}
\label{sec:lcv}

Here, we covariantly generalize typical conservative Newtonian forces. As discussed in the previous section, we restrict ourselves to flat space-time, so the expressions obtained here will be Lorentz covariant, not general covariant. It is also important to note that all cases considered here should only be physically accurate for a force applier in uniform motion,
\be
\label{eq:unif}
U^\mu = \text{constant}.
\ee
As mentioned in section \ref{sec:intro}, an accelerated applier may only communicate its momentum to the test-particle after a finite time interval, if both objects are a finite distance apart. In such a case a field description for $\F^{\mu \nu}$, as in electrodynamics, should be the most appropriate. Nonetheless, even if the applier is initially in uniform motion, there should still be some back-reaction due to its action on the test-particle. We assume it to be negligible by e.g. taking the applier to be very massive compared to the test-particle. In this way eq. \eqref{eq:unif} is dynamically preserved.\footnote{Also note that in curved space-time, condition \eqref{eq:unif}, generalizes to the force applier being in free-fall via the equivalence principle. This is made use of to study the free-falling spring in section \ref{sec:Hooke}.}
\par
Assuming \eqref{eq:unif} also allows for a quantity to be conserved.\footnote{Thus, technically, these are \emph{conservative} forces only if the applier is uniform motion \eqref{eq:unif}.} In the applier's rest frame, this covariant quantity corresponds to the total (kinetic + potential) energy. This is shown explicitly in appendix \ref{app:lagrangian} where the results of this section are re-derived from a Lagrangian approach.
\par

\subsubsection{Coulomb force}
\label{sec:Coulomb}
Let us first consider the Coulomb force because it provides a quick check of \eqref{eq:2F}. Coulomb's law reads
\be
\label{eq:CN}
\vec{F}_N = q{\vec{x}  \over |\vec{x}|^3},
\ee
where $q$ is a constant, $\vec{x}$ is the position of the test-particle with respect to the point charge at the origin. We identify the applier as the point charge which is at rest at the origin.
\par
In the applier's rest frame, where $U^\mu = (1,\vec{0})$, we have
\be
|\vec{x}|^2 = x^2 +  (U \cdot x)^2.
\ee
We can now apply \eqref{eq:presc} to find
\be
F^\mu = q{x^\mu \over \big[x^2 +  (U \cdot x)^2 \big]^{3 \over 2}}
\ee
 and plug into \eqref{eq:2F} to get
\be
\label{eq:C}
\F^{\mu \nu} = q { U^\mu x^\nu - U^\nu x^\mu \over \big[x^2 +  (U \cdot x)^2 \big]^{3 \over 2}},
\ee
which is the electromagnetic field tensor of a point charge $q$ in uniform motion with 4-velocity $U^\mu$ \cite{jackson_classical_1999}. Indeed, we should only expect \eqref{eq:C} to be valid for constant $U^\mu$ since Couloumb's law \eqref{eq:CN} is only valid if the charge is strictly at rest \cite{jackson_classical_1999}.
\par
Note that $x^\mu$ in \eqref{eq:C} should actually be the coordinate difference between test-particle and the charge. In this case we shift
\be
x^\mu(\tau) \to x^\mu(\tau) - X^\mu(\tilde{\tau})
\ee
where $\tilde{\tau}$ is the charge proper time and $X^\mu$ its 4-position. Since the charge is in uniform motion we must have
\be
X^\mu(\tilde{\tau}) = \bar{X}^\mu + \tilde{\tau} U^\mu
\ee
where $\bar{X}^\mu$ is a constant 4-vector specifying the initial 4-position of the charge. Conveniently, any term $x^\mu \propto U^\mu$ does not contribute to \eqref{eq:C} so we may simply shift
\be
x^\mu(\tau) \to x^\mu(\tau) - \bar{X}^\mu
\ee
in \eqref{eq:C} to account for a displaced charge.

\subsubsection{Hooke's law}
\label{sec:Hooke}
Here we covariantly generalize Hooke's law. We consider the force applier as the opposite end of the elastic spring to where the test-particle is attached. We let the applier be at rest at the origin. For a zero rest length spring we have
\be
\label{eq:HN}
\vec{F}_N = - k\vec{x},
\ee
 with $\vec{x}$ the position of the test-particle. From eqs. \eqref{eq:presc} and \eqref{eq:HN} we have
\be
F^\mu = - k x^\mu,
\ee
and \eqref{eq:2F} yields
\be
\label{eq:H}
\F^{\mu \nu} = k (x^\mu U^\nu - x^\nu U^\mu).
\ee
\par
We can now plug \eqref{eq:2F} into \eqref{eq:law} and solve for the motion. Naturally, if the applier is at rest $U^\mu = (1, \vec{0})$ we (trivially) recover the relativistic harmonic oscillator \cite{1957AmJPh..25..535M,1965Natur.205..892H,PhysRevD.6.1474,1994AmJPh..62..531M}. A non-trivial check of \eqref{eq:H} can be made by comparing with the work of Gron \cite{doi:10.1119/1.12623} (see also \cite{ doi:10.1119/1.13978}), where the stress on an elastic body in different inertial frames was studied. Without loss of generality, we let the spring move with velocity $v$ along $x$. This amounts to having both ends of the spring moving solidarily with the following instantaneous parametrization for their 4-velocities,
\be
u^\mu = U^\mu = \gamma(1, v, 0, 0), \qquad \gamma = 1/\sqrt{1 - v^2}.
\ee
Plugging the above into \eqref{eq:H} we get the following for the spatial part of the equations of motion \eqref{eq:law},
\begin{align}
\label{eq:Gron}
{d p_x \over d t} &= - k \gamma (x - v t), \\
\label{eq:Gron2}
{d p_y \over d t} &= - {k \over \gamma}  y, \qquad {d p_z \over d t} = - {k \over \gamma}  z,
\end{align}
where we made use of $dt = \gamma d \tau$. In agreement with Gron \cite{doi:10.1119/1.12623}, we see that there is an effective spring coupling of $k \gamma$ for strains along the direction of motion and of $k / \gamma$ for tranverse strains. 
\par
\par
Finally, we may also make an electromagnetic analogy with Hooke's law. Plugging \eqref{eq:H} into Maxwell's equations \cite{jackson_classical_1999},
\be
\partial_\mu \F^{\mu \nu} = J^\nu,
\ee
gives
\be
J^\mu = 3 k U^\mu,
\ee
which is the 4-current of a homogeneous medium with constant proper volume charge density $3k$ moving with constant 4-velocity $U^\mu$. 
\par

\subsubsection{Constant force}

Arguably, the simplest force in Newtonian mechanics is the constant force,
\be
\label{eq:cte}
\vec{F}_N = \text{const}.
\ee
Plugging \eqref{eq:cte} into Newton's law \eqref{eq:Fma} yields so-called hyperbolic motion \cite{1909AnP...335....1B,1910AnP...338..649S,Misner:1973prb}. 
\par
In our covariant description we require identification of the force applier. To make progress, we note that a uniform electric field is generated by a homogeneously surface charged infinite flat sheet. We may thus identify the sheet as the force applier of a constant force, whose covariant generalization \eqref{eq:2F} should correspond to its electromagnetic field tensor, similarly to the Coulomb force. Hence, we write
\be
\label{eq:plate}
\vec{F}_N = {\sigma \over 2} \vec{n} = \text{const},
\ee
where $\sigma$ is a constant that plays the role of the surface charge density and $\vec{n}$ is the unit normal to the sheet that points in the direction of observation (where the test-particle is). Then, eq. \eqref{eq:presc} yields
\be
F^\mu = {\sigma \over 2} n^\mu
\ee
with\footnote{As mentioned in footnote \ref{fn}, the choice of  $n^0$ on the local Lorentz frame \eqref{eq:normal} is arbitrary. We set it to zero.}
\be
n^\mu|_{\vec{U} = 0}= (0, \vec{n}).
\ee
The above translates into the following covariant constraints on $n^\mu$,
\be
\label{eq:n}
n^2 = 1, \qquad n \cdot U = 0.
\ee
This leaves twofold freedom for $n^\mu$ which are precisely the angles that specify the orientation of the sheet. Eq. \eqref{eq:2F} then reads
\be
\label{eq:cteC}
\F^{\mu \nu} = {\sigma \over 2}(U^\mu n^\nu - U^\nu n^\mu),
\ee
with $n^\mu$ obeying \eqref{eq:n}.
\par
One can now check that \eqref{eq:cteC} is consistent with the electromagnetic field tensor produced by an infinite homogeneously surface charged flat sheet in uniform motion with 4-velocity $U^\mu$ \cite{jackson_classical_1999}.
\par
It is not surprising that a constant force can be written covariantly in terms of a constant `electromagnetic field' $\F^{\mu \nu}$. In fact, \emph{any} constant $\F^{\mu \nu}$ will lead to uniformly accelerated motion \cite{friedman2012making}, which is not surprising. A constant magnetic field also leads to constant, albeit centripetal, acceleration. This analysis is restricted to flat space-time. Indeed, note that eq. \eqref{eq:cteC} is only \emph{Lorentz} covariant: We are assuming that there is a single tangent space at every point on the sheet to which $n^\mu$ belongs, which only holds true in flat space. In section \ref{sec:vms} we show how the variable-mass-rocket can provide a general covariant definition of uniformly accelerated motion.

\section{ Dissipative forces}
\label{sec:cf}

Here, we consider the force due to relative motion of the test-particle with respect to a medium. We find it easiest to compute the force in test-particle's instantaneous rest-frame. Hence, we let $U^\mu = u^\mu$ in \eqref{eq:2F} and reserve the symbol $U^\mu$ for the 4-velocity of the medium at the test-particle's instantaneous position.\footnote{In this case the 4-force $f^\mu = \F^{\mu \nu} u_\nu$ is related to $F^\mu$ via an orthogonal projection $f^\mu = P^{\mu\nu}(u^\alpha) \, F_\nu$ where $P_{\mu \nu}(u^\alpha) \equiv g_{\mu \nu} + u_\mu u_\nu$. The reader familiar with relativistic hydrodynamics may recognize $P^{\mu \nu}(U^\alpha)$, which projects orthogonally to the fluid's 4-velocity $U^\mu$. The projector $P^{\mu \nu}(U^\alpha)$ is a key player in Eckart's covariant formulation of relativistic viscous fluids \cite{1940PhRv...58..919E} (see \cite{rezzolla2013relativistic} for a more recent account).}
\par
Most dissipative forces, in the rest-frame of the test-particle, can be written generically
\be
\label{eq:Ndrag2}
\vec{F}_N = b \, \vec{U},
\ee
for some drag coefficient $b$. Eq. \eqref{eq:Ndrag2} is then covariantly generalized to
\be
F^\mu = B U^\mu,
\ee
with $B$ a covariant drag coefficient, that should depend on invariants built out of $u^\mu$ and $U^\mu$. Due to the mass-shell conditions \eqref{eq:mass-shell} and \eqref{eq:msU} there is only one non-trivial possiblity,
\be
\label{eq:BuU}
B = B(- u \cdot U).
\ee
Then, eq. \eqref{eq:presc} relates the drag coefficients,
\be
\label{eq:B2}
B( U^0) = b.
\ee
Finally, eq. \eqref{eq:2F} gives
\be
\label{eq:2Fc}
\F^{\mu \nu} = B\left(u^\mu U^\nu - u^\nu U^\mu \right),
\ee
which is antisymmetric under interchange of test particle and force applier $u^\mu \leftrightarrow U^\mu$, a resemblance of Newton's third law.\footnote{Note that this is not the case for any of the conservative examples of section \ref{sec:lcv}. In fact, Newton's third law does not hold in a relativistic theory given that interchange of momenta between two bodies cannot be instantaneous, unless the interaction is via direct contact, which is typically the working assumption for most dissipative forces.} 
\par
Note that the form \eqref{eq:2Fc} for $\F^{\mu \nu}$ also follows from  letting $U^\mu$ be the medium 4-velocity in eq. \eqref{eq:2F} and parametrizing instead $F^\mu = - B u^\mu$.\footnote{As done in eq. \eqref{eq:HD} in section \ref{sec:intro}} That is, interpreting the force applier as the fluid which applies a force $\vec{F}_N = - B \vec{u}$ in its rest-frame.
\par
As a simple test consider flat space and take both test-particle and fluid moving colinearly along $x$, 
\begin{align}
\label{eq:up}
u^\mu &= \gamma(1,v,0,0), \qquad \gamma = 1/ \sqrt{1 - v^2}, \\
U^\mu &= \Gamma(1,V,0,0), \qquad \Gamma = 1/ \sqrt{1 - V^2}.
\end{align}
Plugging the above parametrizations into eq. \eqref{eq:2Fc}, and using $dt = \gamma d\tau$, the $x$ component of \eqref{eq:law} reads
\be
\label{eq:ldN}
{d (\gamma v) \over d t} = - B\big(\gamma \Gamma (1  -v V)\big) \, {\gamma \Gamma} (v - V),
\ee
where we made explicit dependence \eqref{eq:BuU}. Eq. \eqref{eq:ldN} takes a more familiar form when use is made of the formula for relativistic addition of velocities \cite{jackson_classical_1999},
\be
\label{eq:vrel}
v' = {v - V \over 1 - v V}, \qquad \gamma' = {1 \over \sqrt{1 - v'^2}}.
\ee
Eq. \eqref{eq:ldN} can then be written as
\be
m {d (\gamma v) \over d t} = - B(\gamma') \, m \gamma' v',
\ee
where we briefly reinstated $m$ for physical clarity. Eq. \eqref{eq:ldN} states the force on the test-particle is proportional, and against, its  momentum \emph{relative}  to the medium, $\vec{p}' = m \gamma' v'$, i.e. according to \eqref{eq:vrel}.
\par
Now, similarly to what was done for the conservative forces in section \ref{sec:lcv} we may take any Newtonian drag force and covariantly generalize it. For example, Stokes' drag \cite{1851TCaPS...9....8S} reads
\be
\label{eq:Stokes}
\vec{F}_N = 6 \pi \mu  R\,  \vec{U}
\ee
where $\mu$ is the fluid viscosity, $R$ is the sphere radius and $\vec{U}$ is, non-relativistically, also the velocity of the fluid. It then follows that the drag coefficient $b = 6 \pi \mu R$ is constant so \eqref{eq:B2} immediately reads
\be
B = 6 \pi \mu R.
\ee
\par
Unfortunately, Stokes' drag, which follows from Newtonian hydrodynamics, should not be valid relativistically.\footnote{In fact, flow should become turbulent way before relativistic velocities are achieved, so Stokes' law \eqref{eq:Stokes} would not apply either way.} This is the case for most  dissipative forces, which are typically derived within a Newtonian setting. Naturally, this does not prevent covariant generalization of these forces, rather, we may covariantly express their domain of validity. Non-relativistic  \emph{relative} motion between test-particle and the medium amounts to having the fluid moving very slowly, i.e. $U^0 \sim 1$,  in the test-particle's rest frame. This condition reads covariantly
\be
\label{eq:nonrelreg}
\text{Non-relativistic regime: } \; (u \cdot U) \sim - 1.
\ee
To go beyond the non-relativistic regime one should re-derive these dissipative forces within a special relativistic setting, e.g. using relativistic hydrodynamics \cite{rezzolla2013relativistic}.
\par
Here we show that for a certain class of dissipative forces the relativistic derivation can be done straightforwardly. Namely, when the size of the test-particle, i.e. the perturber inside the medium, is of the order, or smaller, than the mean free path of a particle in the medium. That is, when the Knusden number $\gtrsim 1$. In this case a continuum hydrodynamical description is not adequate. Instead, kinetic theory applies, where one may consider the drag as a result of numerous individual scattering events between the test-particle and each medium constituent. 
\par
In what follows we will compute $B$ for a variety of well known models. We first consider, in section \ref{ssec:scat}, the drag force due to scattering in a medium, where no change to the test-particle's mass is undergone. In section \ref{sec:vms} we consider variable-mass-effects, in particular the force due to accretion and the variable-mass-rocket.

\subsection{Drag due to collisions in a  medium}
\label{ssec:scat}

Consider the motion of a test-particle through a medium, a field of point-like objects of much smaller mass $M$. Each medium constituent will have momentum
\be
P^\mu = M U^\mu
\ee
with $M \ll m = 1$ and $U^\mu$ the 4-velocity of an individual constituent.\footnote{We regret the slightly misleading notation for $M$ as the mass of an individual constituent.} 
\par
 The particles of the medium will then-scatter on the test-particle and have an overall dissipative effect on its motion. To estimate this effect we work on the comoving frame of the test particle. If the medium has proper particle density $n$, the test particle will see a Lorentz contracted density $\gamma n$, where
\be
U^\mu = \gamma (1, \vec{v}), \qquad \gamma = 1/\sqrt{1 - v^2}.
\ee
The medium constituents will thus scatter off the test-particle at a rate
\be
\label{eq:dnu}
d \nu = n |\vec{U}| \, d \sigma
\ee
where $|\vec{U}| = \gamma v$ and $d \sigma$ is the differential cross-section of the interaction between the test-particle and an individual constituent.
\par
After scattering, there will be a momentum shift on each constituent. Momentum conservation then implies a momentum shift on the test-particle itself, which will naturally depend on the scattering angle. When multiplied by the collision rate \eqref{eq:dnu} the momentum shift gives the infinitesimal force on the test-particle. The total force on the test-particle, due to scattering in the medium, is thus
\be
\label{eq:scatf}
\vec{F}_N = \int \! \Delta \vec{p} \, d \nu = \int n \, |\vec{U}|  \left({d \sigma \over d \Omega}\right)  \Delta \vec{p}\,(\vec{U}, \Omega) \, d\Omega
\ee
where $\Delta \vec{p}$ the momentum shift on the test-particle, and $d \Omega$ is the solid angle element.
\par
If the scattering process is elastic, i.e. if each medium constituent exits the scattering event with the same energy as they entered with, then the scattering angle fully specifies the momentum shift. This can be seen in the following way. We decompose the momentum shift into parallel and orthogonal components to $\vec{U}$,
\be
\Delta \vec{p} =  \Delta \vec{p_\|} + \Delta \vec{p_\perp}.
\ee
If the differential cross-section only depends on the polar angle $\theta$, then the orthogonal component $\Delta \vec{p_\perp}$ integrates to zero in eq. \eqref{eq:scatf}. The force will then be parallel to the medium velocity $\vec{U}$, and will have the form \eqref{eq:Ndrag2}. In terms of the scattering angle $\theta$, the parallel component reads
\be
\label{eq:el}
\Delta \vec{p}_\text{el.} =   \Delta \vec{p_\|} = M \vec{U} (1 - \cos \theta).
\ee
\par
If, instead, the scattering process is inelastic, in the sense that particles deposit their full momentum by e.g. ``sticking" to the test particle after colliding, then the momentum shift will be simply given by the initial momentum,
\be
\label{eq:inel}
\Delta \vec{p}_\text{inel.} = M \vec{U}.
\ee
\par
We assume here that no increase in the test-particle's rest mass is undergone. The extra mass is randomly diffused away by evaporation or some similar process \cite{PhysRev.23.710}. Importantly, we do not need to specify the mechanism, only that the test-particle's mass remains unchanged. In this case, the integral \eqref{eq:scatf} is trivial and is given by
\be
\label{eq:dragN}
\vec{F}_\text{inel.}  = \rho \sigma |\vec{U}| \vec{U}
\ee
with $\rho = M n$, the proper mass density of the medium, and $\sigma$ the total scattering cross section (which, depending on the nature of the interaction, may still be a function of $|\vec{U}|$).

\subsubsection{Dust}

\emph{Hard-sphere scattering.} Consider the test-particle as a sphere of radius $R$. The hard-sphere differential scattering cross section is constant and reads \cite{huang2000statistical}
\be
\label{eq:hs}
{d \sigma \over d \Omega} = { R^2 \over 4} 
\ee
Considering the collision to be elastic the dust particles will specularly reflect off the test particle. Plugging \eqref{eq:el} into \eqref{eq:scatf} and integrating over the angles gives back eq. \eqref{eq:dragN} with 
\be
\sigma = \pi R^2.
\ee
The fact that the elastic and inelastic models yield the same result is a peculiarity of the spherical shape of the test particle.
\par
Comparing \eqref{eq:dragN} with \eqref{eq:Ndrag2} we read off
\be
b = \rho \sigma |\vec{U}| = \rho \sigma \sqrt{(U^0)^2 - 1}.
\ee
From \eqref{eq:B2} and \eqref{eq:BuU} we then find the covariant drag coefficient due to hard-sphere collisions in a dust medium,
\be
\label{eq:Bdust}
B = \rho \sigma \sqrt{(u \cdot U)^2 - 1}.
\ee
\par
Naturally, in many situations the purely elastic/inelastic models for the momentum shift are inadequate. For example, in the ultrarelativistic motion of dust grains through interstellar dust, impinging ions have a penetration length far greater than the typical diameter of dust grains, meaning that they only leave a fraction of their momentum on the test particle \cite{Hoang:2017ijx}.
\par
There is also the possibility that dust grains themselves become ionized after scattering \cite{1979ApJ...231...77D}, medium particles reflect diffusely off the test particle (instead of specularly) \cite{PhysRev.23.710}, or quantum diffractive effects become relevant \cite{2005}.
\par

\emph{Gravitational scattering.} We now let the test particle have enough mass to be a gravtional perturber. When drifting across a field of matter, it will gravitationally deflect the surrounding matter, which then must backreact on the perturber itself. On average this results in a force opposing the perturber's velocity, a dissipative effect known as \emph{dynamical friction}, first studied in detail by Chandrasekhar \cite{Chandra}.
\par
Naturally, dynamical friction is not a typical drag force since it is due to a long-distance interaction. However, if the perturber's mass is much smaller than the curvature scale of the background metric, as in EMRIs where the curvature scale is set by the largest black hole mass, one can still see dynamical friction as a local efffect. Note that a similar assumption occurs in Newtonian computations of dynamical friction, where it is assumed that in the victinity of the test-particle one may just consider its own gravitational field.
\par
To estimate the dynamical friction effect we can make use of the gravitational scattering cross-section in \eqref{eq:scatf}. We let the perturber be a Schwarzschild black hole. At leading order in Newton's constant $G$, the differential cross-section, in the perturber's rest frame, is \cite{Collins:1973xf,Doran:2001ag}
\be
\label{eq:gcs}
{d \sigma \over d \Omega} = {G^2 (1 + v^2)^2  \over 4 v^4 \sin^4(\theta / 2)},
\ee
where $v$ is the dust particle initial velocity, and $\theta$ the deflection angle. The non-relativistic limit $v \to 0$ yields the Rutherford formula.
\par
Since at leading order in $G$ the scattering process is conservative, we make use of the elastic momentum shift \eqref{eq:el} in \eqref{eq:scatf}. Plugging \eqref{eq:gcs} into \eqref{eq:scatf}, and making use of
\be
U^\mu = \gamma (1, \vec{v}), \qquad \gamma= 1/ \sqrt{1 - v^2},
\ee
we find
\be
\label{eq:dynf}
\vec{F}_N =  4 \pi \rho \, G^2  \Lambda {(1+v^2)^2  \over   (1 - v^2) }\, {\vec{v} \over  v^3},
\ee
where $\Lambda \equiv \ln \left({ b_\text{max} \over b_\text{min}} \right)$ is the Coulomb logarithm. The maximum impact parameter $b_\text{max}$ is set by the size of the matter field, while the minimum impact parameter $ b_\text{min}$ is determined by the effective size of the perturber, the largest of either the perturber physical size or the capture impact parameter. In the latter case $b_\text{min}$ will depend on $v$, and for impact parameters smaller than $b_\text{min}$ accretion will occur (see section \ref{sec:vms}).
\par
Expression \eqref{eq:dynf} was first obtained in \cite{Petrich} and more recently in the weak field limit of \cite{Vicente:2022ivh}. 
Compared to the Newtonian result for a dust medium \cite{Chandra}, eq. \eqref{eq:dynf} is corrected by the factor $(1+v^2)^2  /  (1 - v^2) = \gamma^2 (1+v^2)^2$. Its origin can be dissected into the $(1+v^2)^2$ relativistic correction to the weak-field gravitational cross-section \eqref{eq:gcs}, a $\gamma$ factor due to Lorentz contraction of the medium density in the perturber's rest frame and a further $\gamma$ factor coming from the (relativistic) momentum shift in \eqref{eq:scatf}.

\par
From \eqref{eq:Ndrag2} we read off $b$ and, from \eqref{eq:B2} and \eqref{eq:BuU}, we find the covariant dynamical friction coefficient
\be
\label{eq:Bdustg}
B = 4 \pi \rho \, G^2 \Lambda { [2 (u \cdot U)^2 - 1]^2  \over   [(u \cdot U)^2 - 1]^{3 \over 2} }, 
\ee
with corresponding 4-force, which follows from contracting \eqref{eq:2Fc} with $u_\nu$,
\be
f^\mu = 4 \pi \rho \, G^2 \Lambda { [2 (u \cdot U)^2 - 1]^2  \over   [(u \cdot U)^2 - 1]^{3 \over 2} } \Big(u^\mu (u \cdot U) + U^\mu \Big).
\ee
This is consistent with the tangential drag considered in \cite{Barausse:2007dy}.
\subsubsection{Radiation} 

 If the medium consists of null particles like photons, and they are scattered by the test-particle, then there should also be a back reaction on the test-particle. However, the argument must be adapted due to the absence of a rest frame for the photons. There is no concept of proper photon density.
\par
Instead, one often has information in the rest frame of the radiation source (e.g. a star).
In this frame, each photon has energy $E$, and we can write the photon momentum as
\be
\label{eq:Pphoton}
P^\mu = E U^\mu
\ee
with
\be
\label{eq:Un}
U^\mu = (1, \vec{n}),
\ee
where $\vec{n}$ is a unit vector that specifies the travel direction of the photons. 
\par
$U^\mu$ is a 4-vector with only 2 degrees of freedom (the angles on a sphere). One of the initially 4 degrees of freedom is removed by the fact that $U^\mu$ is a null vector. The other is removed by the normalization $U^0 = 1$ in the rest frame of the source. If $u_S$ is the four-velocity of the source, these constraints can be written covariantly as
\be
\label{eq:U2uS}
U^2 = 0, \qquad u_S  \cdot U = - 1.
\ee
It is easy to check that in the source rest-frame $u_S^\mu = (1, \vec{0})$, the above restricts $U^\mu$ to be of the form \eqref{eq:Un}.
\par
Now, in the particle's instantaneous rest frame, $u^\mu = (1, \vec{0})$, we parametrize
\begin{align}
\label{eq:usU}
u_S^\mu &= \gamma_S(1,\vec{v}_S), \qquad \gamma_S = 1/ \sqrt{1 - v_S^2}, \\
U^\mu &= (U^0, \vec{U}).
\end{align}
If, for example, we let the photons move colinearly with the source, $\vec{U} \cdot \vec{v}_S = |\vec{U}| v_S$, conditions \eqref{eq:U2uS} then fix
\be
\label{eq:blueshift}
U^0 = |\vec{U}| = \sqrt{1 + v_S \over 1 - v_S},
\ee
which is the relativistic longitudinal Doppler factor corresponding to a blueshift. Indeed, according to eq. \eqref{eq:blueshift} the energy and momentum of the photon \eqref{eq:Pphoton} will be blueshifted with respect to the energy $E$ it is originally emitted with by the source (i.e. in its rest-frame). This is expected given that the test-particle sees the source moving towards it with speed $v_S$.
\par
In fact, regardless of the relative orientation of the velocity $\vec{v}_S$ of the source and the radiation direction $\vec{U}$, in the test-particle's frame, $U^0 = |\vec{U}|$ will always be the Doppler shift factor. This simply follows from repeating the previous argument with generic $\vec{U}$ and $\vec{v}_S$.\footnote{In this case we find $U^0 = 1/ \gamma_S (1 - v_S \cos \theta)$ where $\theta$ is the angle between $\vec{v}_S$ and $\vec{U}$.}
\par
Therefore, in conclusion, we find that the collision rate, in the test-particle's frame, will still be given by eq. \eqref{eq:dnu} where now $n$ is the photon number density in the rest frame of the source and the $|\vec{U}|$ factor accounts for the Doppler shift of the collision rate.\footnote{This follows from a typical Doppler analysis: If every $\Delta t_S$ seconds a photon is produced by the source, then consecutive photons will be separated by a distance $d = c \Delta t_S$, in the source rest-frame. If the test-particle is moving towards the source at speed $v$, the time $\Delta t$ that elapses between two photons being received by the test-particle follows from $c \Delta t + v \Delta t = d / \gamma$, where $\gamma = 1/\sqrt{1-v^2}$ accounts for the Lorentz contraction of the distance $d$ in the test-particle's rest frame. We then find $\Delta t = \sqrt{1 - v \over 1 + v} \Delta t_S$ or $\Delta t^{-1} = |\vec{U}| \Delta t^{-1}_S$.} Eq. \eqref{eq:scatf} will also retain the same form, where $\Delta \vec{p}$ is (minus) the momentum shift of an individual photon. Finally, the formulas for the elastic and inelastic momentum shift \eqref{eq:el} and \eqref{eq:inel} will also hold provided $M$ is replaced by $E$,
\begin{align}
\label{eq:el2}
\Delta \vec{p}_\text{el.} &= E \vec{U} (1 - \cos \theta). \\
\Delta \vec{p}_\text{inel.} &= E \vec{U}. \label{eq:inel2}
\end{align}
The only difference lies in the fact that $U^\mu$ is constrained by \eqref{eq:U2uS}, instead of being time-like, as in the massive case.
\par
\emph{Hard-sphere scattering.}
As with the dust case, we first consider the test-particle to be a hard-sphere of radius $R$ with differential cross-section given by eq. \eqref{eq:hs}. Either of the elastic and inelastic models \eqref{eq:el2} and \eqref{eq:inel2} integrate in \eqref{eq:scatf} to 
\be
\vec{F}_N = e \sigma |\vec{U}| \vec{U}
\ee
with $\sigma = \pi R^2$ and $e \equiv n E$, the energy density in the source rest frame.\footnote{For the reader familiar with the literature on the Poynting-Robertson effect note that $e = I/c$, where $c$ is the speed of light (assumed $c=1$ in this work) and $I$ is the radiation energy flux density, also sometimes known as \emph{intensity}.}
\par
We then read off, from eq. \eqref{eq:Ndrag2},
\be
b = e \sigma |\vec{U}| = e \sigma U^0
\ee
which, from \eqref{eq:B2} and \eqref{eq:BuU}, yields the covariant coefficient
\be
\label{eq:Brp}
B = - e \sigma (u \cdot U).
\ee
This result matches the $(u \cdot U)^2 \to \infty$ limit of the dust drag coefficient \eqref{eq:Bdust} (recall that $u \cdot U < 0$).
\par
The Poynting-Robertson 4-force \cite{1937MNRAS..97..423R} follows from contracting \eqref{eq:2Fc} with $u_\nu$,
\be
\label{eq:RP}
f^\mu = - e \sigma (u \cdot U) \left[ (u \cdot U) u^\mu + U^\mu \right].
\ee
\par
Note that due to the term proportional to $u^\mu$, the force \eqref{eq:RP} will always have a term opposing the particle's velocity, even if photons flow orthogonally to the particle's velocity. This would occur for example in a circular orbit of a test-particle around a star, a radially emitting source.
\par
To be concrete we may choose the following parametrization on the star's rest frame, for which radiation is being emitted in the $y$ direction and the particle is moving along $x$,
\begin{align}
\label{eq:usU2}
u^\mu &= \gamma(1,v,0,0), \qquad \gamma = 1/ \sqrt{1 - v^2}, \\
U^\mu &= (1, 0, 1, 0).
\end{align}
Then, making use of $dt = \gamma d\tau$, we get for the force components,
\be
{dp_x \over dt} = - e \sigma \gamma^2 v, \qquad {dp_y \over dt} = e \sigma.
\ee
The component along $y$ is the expected radiation pressure force, while the component along $x$ is the Poynting-Robertson drag.
\par
As with the dust drag case, here we have assumed the elastic and inelastic models \eqref{eq:el2}, amounting  to perfect (specular) reflection and absorption/emission,\footnote{Again we emphasize that we need not consider any particular thermodynamical model for the emission. Even though the photon momentum is completely absorbed by the test-particle we do not allow for its rest-mass $m$ to increase. We make, however, the relatively safe assumption that there exists some emission process that allows for this (isotropic emission of thermal radiation is one possibility).} respectively. In reality this is mostly not case and the drag coefficient \eqref{eq:Brp} should be multiplied by some ``efficiency factor", which may be determined by a microscopic model of the interaction between radiation and the precise chemical composition of the test-particle \cite{1979Icar...40....1B}.
\par
\emph{Gravitational scattering.} As for the dust case we assume the test particle to be a sufficiently heavy Schwarzschild black hole to affect a massless medium. Scattered photons will follow null geodesics in a Schwazschild space-time. To lowest order in $G$ the deflection angle can be computed, and from there the scattering cross-section.
\par
Alternatively, the result can be directly obtained by taking the limit $v \to 1$ in eq. \eqref{eq:gcs},
\be
{d \sigma \over d \Omega} = {G^2  \over \sin^4(\theta / 2)}.
\ee
Plugging the above into \eqref{eq:scatf} and choosing the elastic momentum shift \eqref{eq:el2} yields
\be
\label{eq:gsp}
\vec{F}_N = 16 \pi e G^2 \Lambda |\vec{U}| \vec{U}.
\ee

To our understanding, this result is consistent with the massless weak-field computation of \cite{Vicente:2022ivh}. It differs, however, by a factor of $4/3$ from the ultra-relativistic expression of \cite{Syer:1994vr} which assumes an isotropic distribution of velocities for the medium (here we consider a collimated flow of photons, i.e. null dust).
\par
From \eqref{eq:gsp} we read off the drag coefficient, $b =  16 \pi e G^2 \Lambda U^0$ which generalizes to
\be
\label{eq:Bgr}
B = - 16 \pi e G^2 \Lambda \, (u \cdot U).
\ee
The covariant dynamical friction coefficient due to radiation has the same functional form as the Robertson-Poynting coefficient \eqref{eq:Brp}. It also matches the ultra-relativistic limit, $(u \cdot U)^2 \to \infty$, of the dynamical friction dust coefficient \eqref{eq:Bdustg}.

\subsubsection{Gas }

\par
One has to be slightly more sophisticated if the medium is collisional, i.e. a gas. Instead of moving in a collimated flow, the particles of the medium will be dispersed in 4-velocity $k^\mu$  (with $k^2 = -1$) according to some distribution function $W(\vec{k})$. The test-particle will see $W(\vec{k}) d^3 \vec{k}$ particles per unit volume with momentum $M \vec{k}$. The differential collision rate is then generalized from eq. \eqref{eq:dnu} to
\be
\label{eq:dnug}
d \nu =  {|\vec{k}| \over k^0} W(\vec{k}) \, d \sigma \, d^3 \vec{k}
\ee
with $k^0 = \sqrt{1 + |\vec{k}|^2}$.
The force on the test particle, due to scattering in the gas, instead of \eqref{eq:scatf} now reads
\be
\label{eq:scatfg}
\vec{F}_N = \int  |\vec{k}|  W(\vec{k}) \left({d \sigma \over d \Omega}\right)  \Delta \vec{p}\,(\vec{k}, \Omega) \, d\Omega \,{ d^3 \vec{k} \over k^0} 
\ee
where $\Delta \vec{p}$ the momentum shift on the test particle, and $d \Omega$ is the solid angle element.
\par
Here we will make use of the Maxwell-Boltzmann distribution,
\be
\label{eq:MB}
W(\vec{k}) =  n \left({ \beta \over 2 \pi } \right)^{3 \over 2}\exp\left[{- {\beta \over 2 }  (\vec{k} - \vec{U})^2 }\right]\!,
\ee
where 
\be
\beta \equiv {M \over k_B T},
\ee
with $M$ the mass of a single gas molecule, $k_B$ the Boltzmann constant and $T$ the gas temperature. Note, however, that $W(\vec{k})$ should be Lorentz invariant in order for $W(\vec{k}) d^3 \vec{k}$ to transform as a number density. This means that \eqref{eq:MB} is only applicable if the gas is non-relativistic. One alternative is to use a relativistic equilibrium distribution \cite{osti_4752122,1963JMP.....4.1163I} such as the Maxwell-Jüttner distribution \cite{1911AnP...339..856J} (see eq.  \eqref{eq:MJ}). It is known, however, that at ultra-relativistic temperatures the Maxwell-Jüttner distribution becomes inadequate due to quantum effects becoming relevant, such as pair production and particle indistinguishability \cite{huang2000statistical}. Nonetheless, most astrophysical gases have non-relativistic temperatures. For example, accretion disks can reach temperatures up to $\sim 10^8$ K \cite{2011hea..book.....L}. Letting $M$ be the mass of the electron, one has $\beta  \sim 10^3 \gg 1$. Having $\beta \gg 1$ makes a gas non-relativistic in its own rest frame. However, eq. \eqref{eq:scatfg} is in the rest frame of the particle, which may observe the gas moving relativistically. The Maxwell-Boltzmann distribution will be valid to a good approximation if the relative motion is non-relativistic as well, $|\vec{U}| \ll 1$. Covariantly, this can be expressed as condition \eqref{eq:nonrelreg}.

\par
\emph{Hard-sphere scattering.} Plugging the hard-sphere scattering cross-section \eqref{eq:hs} into \eqref{eq:scatfg} and making use of either the elastic or inelastic models \eqref{eq:el} and \eqref{eq:inel} we get
\be
\label{eq:hsgas}
\vec{F}_N = M\sigma \int \! \vec{k} \, |\vec{k}|  \, W(\vec{k})  \,{ d^3 \vec{k} \over k^0},
\ee
with $\sigma = \pi R^2$. Essentially, this is an average of the dust expression \eqref{eq:dragN} over the momentum distribution $W(\vec{k})$. Plugging \eqref{eq:MB} leads to the Newtonian result \cite{shen2006rarefied},\footnote{Since the integrand has support over non-relativistic values of $\vec{k}$ we let $k^0 \to 1$ in \eqref{eq:hsgas} and \eqref{eq:ggas}. \label{k0}}
\begin{align}
\label{eq:fNhs}
\vec{F}_N &= {\rho \sigma \vec{U} \over \beta^2 |\vec{U}|^3} \bigg[ \sqrt{2 \beta \over \pi}  |\vec{U}| (1 + \beta |\vec{U}|^2) \, e^{- {\beta |\vec{U}|^2 \over 2}} 
\notag \\ & \;\;\;\;\;\;+ \erf\bigg(\sqrt{\beta \over 2}|\vec{U}| \bigg)  ( \beta^2 |\vec{U}|^4 + 2  \beta |\vec{U}|^2 - 1  ) \bigg],
\end{align}
where `$\erf$' is the error function. From \eqref{eq:fNhs} we find the covariant drag coefficient
\begin{align}
\label{eq:bNhs}
B &= {\rho \sigma  \over \beta^2 X^3} \bigg[ \sqrt{2 \beta \over \pi}  X (1 + \beta X^2) \, e^{- {\beta X^2 \over 2}} 
\notag \\ & \;\;\;\;\;\;+ \erf\bigg(\sqrt{\beta \over 2} X \bigg)  ( \beta^2 X^4 + 2  \beta X^2 - 1  ) \bigg],
\end{align}
with 
\be
\label{eq:X}
X \equiv \sqrt{(u \cdot U)^2 - 1}.
\ee
Expression \eqref{eq:bNhs} is valid in the non-relativistic regime $X \ll 1$ and $\beta \gg 1$.
\par
For a slow moving gas compared to its thermal agitation, where $X \ll 1/ \sqrt{\beta}$, also known as Epstein regime \cite{PhysRev.23.710}, expression \eqref{eq:bNhs} reduces to the constant Epstein coefficient
\be
\label{eq:BEpstein}
 B = {4 \over 3} \rho \sigma \sqrt{8 \over \pi \beta}.
\ee
\par
In the zero-temperature limit $\beta \to 0$, the gas becomes pressureless and the covariant drag coefficent \eqref{eq:bNhs} goes to the dust expression \eqref{eq:Bdust}. Note that the fully relativistic result is recovered, while \eqref{eq:bNhs} is only valid non-relativistically. This may be traced back to the fact that the hard-sphere differential cross-section \eqref{eq:hs} has the same form in both regimes. The same does not happen for gravitational scattering (see below).
\par
In appendix \ref{app:relgas} we repeat the computation of \eqref{eq:bNhs} using instead the Maxwell-Jüttner distribution with arbitrary relativistic momentum $|\vec{U}|$ in the saddle-point approximation $\beta \to \infty$ (corresponding to a non-relativistic gas in its rest frame, but with arbitrary average speed).
\par
\par
\emph{Gravitational scattering.}
Making use of the gravitational scattering cross-section \eqref{eq:gcs} and the elastic momentum transfer \eqref{eq:el} we get
\be
\label{eq:ggas}
\vec{F}_N = 4 \pi G^2 M \Lambda \int \!  \left[2 (k^0)^2 - 1 \right]^2  { \vec{k}  \over |\vec{k}|^3}  \, W(\vec{k})  \,{ d^3 \vec{k} \over k^0},
\ee
which convolutes the dust dynamical friction \eqref{eq:dynf} over the distribution $W(\vec{k})$. This is a relativistic version of Chandrasekhar's dynamical friction \cite{Chandra} over a generic relativistic $W(\vec{k})$ momentum distribution.
\par
Plugging, the Maxwell-Boltzmann distribution \eqref{eq:MB}, leads however back to Chandrasekhar's expression \cite{Chandra},\footref{k0}
\begin{align}
\vec{F}_N = \frac{4 \pi \rho G^2 \Lambda \vec{U}}{|\vec{U}|^3}    \left(\erf\bigg(\sqrt{\beta \over 2}|\vec{U}| \bigg)- \sqrt{\frac{ 2\beta }{\pi }} |\vec{U}| e^{-\frac{\beta  |\vec{U}|^2}{2}}\right),
\end{align}
\par
from which we read off
\be
\label{eq:Bgasg}
B =  \frac{4 \pi \rho G^2 \Lambda }{X^3}    \left(\erf\bigg(\sqrt{\beta \over 2} X \bigg)- \sqrt{\frac{ 2\beta }{\pi }} X e^{-\frac{\beta X^2}{2}}\right),
\ee
with $X$ given by eq. \eqref{eq:X}. As with the previous case, this expression is valid in the non-relativistic regime $X \ll 1$ and $\beta \gg 1$.
\par
In the slow gas regime (compared to the thermal speed), i.e. where $X \ll 1/ \sqrt{\beta}$, expression \eqref{eq:Bgasg} reduces to the constant coefficient
\be
\label{eq:Bgasge}
B  = \frac{4}{3} \sqrt{2 \pi }  \rho \, G^2 \Lambda \beta ^{3/2}.
\ee
\par
In the zero-temperature limit $\beta \to 0$, the gas becomes pressureless and the covariant drag coefficent \eqref{eq:Bgasg} goes to the dynamical friction dust expression \eqref{eq:Bdustg} (in the non-relativistic regime \eqref{eq:nonrelreg}).

\subsection{Variable-mass systems}
\label{sec:vms}
We now allow for the test-particle to accelerate (deccelerate) due to mass loss (gain). Now we can no longer set the test particle's mass  $m = 1$. We may reinstate the mass by multiplying the LHS of the equations of motion \eqref{eq:law} by $m$. Instead, we reabsorb $m$ into $F^{\mu \nu}$ by redefining
\be
\label{eq:Fm}
F^{\mu \nu} \to {F^{\mu \nu} \over m}.
\ee
\par
We work in the instantaneous rest frame of the test particle where we let it capture (eject) a particle of mass $M \ll m$ with 4-velocity $U^\mu = (U^0,\vec{U})$. Energy conservation implies that the test particle's mass will change by
\be
\label{eq:vme}
dm = \pm   M U^0.
\ee
where $(+)$ is for capture and $( - )$ for ejection. Momentum conservation then requires that the test-particle will get a velocity shift $d \vec{v}$ given by
\be
\label{eq:vmp}
m \, d\vec{v} = \pm M \vec{U}.
\ee
Making use of \eqref{eq:vme} in \eqref{eq:vmp} and dividing by $d \tau$ we get
\be
\label{eq:fvmr}
\dot{\vec{v}} = {\dot{m} \over m} {\vec{U} \over U^0}.
\ee
Given that $\gamma = 1/\sqrt{1 - v^2} = 1$ in the instantaneous rest-frame, we see that the above is in the form of eq. \eqref{eq:Ndrag2} with 
\be
b = {\dot{m} \over m} {1 \over U^0},
\ee
and covariant coefficient
\be
\label{eq:Bvmr}
B = - {\dot{m} \over m} (u \cdot U)^{-1},
\ee
which is valid both for mass capture and ejection, depending on the sign of $\dot{m}$.

\subsubsection*{Variable-mass rocket}
The variable-mass rocket propels itself by ejecting part of its mass (propellant). For the variable-mass rocket one usually has information on the rocket's co-moving frame. Namely, the mass depletion rate $\dot{m}/m$, the rate at which the rocket loses mass, and the exhaust velocity $v_e$, the velocity at which the propellant exits the rocket, both measured by instruments co-moving with the rocket.
\par
Now, the exhast velocity $v_e$ constrains the form of $U^\mu$. Noting that $|\vec{U}| / U^0 = v_e$ on the co-moving frame, we can write this covariantly as
\be
\label{eq:ve}
v_e^2 = {(u \cdot U)^2 - 1 \over (u \cdot U)^2}.
\ee
There is still two-fold freedom in the choice of $U^\mu$ corresponding to the direction of propellant ejection.
\par
We may now plug \eqref{eq:Bvmr} into $\F^{\mu \nu}$ given in eq. \eqref{eq:2Fc} and then contract with $u_\nu$ to get the 4-acceleration,
\be
\label{eq:avmr}
a^\mu = \F^{\mu \nu} u_\nu =  - {\dot{m} \over m} \left(u^\mu +{U^\mu \over u \cdot U} \right)
\ee
in a generic frame.
\par
Choosing the metric to be Minkowski and letting $\vec{u}$ and $\vec{U}$ be collinear leads to the relativistic rocket equation \cite{forward1995transparent}.
\par
Also note that the proper acceleration $a^2 = a_\mu a^\mu$ is uniquely determined by the depletion rate and the exhaust velocity, 
\be
\label{eq:propera}
a^2 = \left({\dot{m} \over m} v_e\right)^2,
\ee
where we made use of eq. \eqref{eq:ve}.
\par
We thus see that if the product of the depletion rate and the exhaust velocity is constant, the rocket will measure a constant acceleration. The rocket will then follow hyperbolic motion \cite{PhysRev.119.2082,Misner:1973prb}, as seen from an outside inertial observer. The force $\F^{\mu \nu}$ on the variable-mass rocket can thus be seen as a ``constant force", in the sense of the proper acceleration \eqref{eq:propera} being constant.
\par
Knowing the value of $a$ one can instead find how the rocket mass decreases by integrating over \eqref{eq:propera}
\be
m(\tau) = m(0) \, e^{- a \tau / v_e}
\ee
in agreement with \cite{Henriques:2011aq}.
\subsubsection*{Accretion}
Accretion is the process via which an object increases its mass by capturing surrounding particles. Knowing the proper accretion rate $\dot{m}/m > 0$ one may directly make use of eq. \eqref{eq:Bvmr}. Now, $B > 0$, indicating that accretion leads to an effective drag force on the test-particle, which is expected given that the test-particle is increasing its inertia. 
\par
We may compute the accretion rate as follows. Eq. \eqref{eq:vme}, with the $(+)$ sign, gives how the mass shifts due to capture of a single particle of mass $M \ll m$. Multiplying by the collision rate,
\be
\label{eq:nuac}
\nu = n \sigma_c |\vec{U}|,
\ee
where $n$ is the proper density of the medium and $\sigma_c$ is the \emph{capture cross section}, we get the accretion rate
\be
\dot{m} = M U^0 \nu = \rho_c \sigma U^0 |\vec{U}|,
\ee
with $\rho = n M$, which reads covariantly
\be
\label{eq:mdotacc}
\dot{m} = - \rho_c \sigma (u \cdot U) \sqrt{(u \cdot U)^2 - 1}.
\ee
The covariant coefficient \eqref{eq:Bvmr} will then read, 
\be
\label{eq:Bacc}
B = {\rho \sigma_c \over m}  \sqrt{(u \cdot U)^2 - 1},
\ee
with $m(\tau)$ evolving according to \eqref{eq:mdotacc}. 
\par
The 4-acceleration on the test-particle also follows from \eqref{eq:avmr}. In terms of the 4-force we have instead
\be
\label{eq:facc}
f^\mu \equiv {D (m u^\mu) \over d \tau} = - {{\dot{m}} \,  \over u \cdot U} U^\mu = \rho \sigma  U^\mu \sqrt{(u \cdot U)^2 - 1}.
\ee
Note that $f \cdot p  \propto f \cdot u \neq 0$ since the rest-mass $m$ of the test-particle is variable. The 4-acceleration $a^\mu$ and 4-velocity $u^\mu$ are still orthogonal, however, since $u^2 = -1$ is preserved.

\emph{Hard-sphere capture.} If the interaction with the medium is via hard-sphere collisions, then the capture cross section $\sigma_c$ can be taken as $\sigma_c = \pi R^2$ with $R$ the radius of the sphere, due to particles of the medium ``sticking" to the sphere after colliding. We see that the accretion drag coefficient \eqref{eq:Bacc} will then match the dust drag coefficient \eqref{eq:Bdust} with the difference that the test-particle's mass $m$ now evolves according to \eqref{eq:mdotacc}.\footnote{Given the finite size of a dust particle one should also expect the test-particle to eventually increase its volume $\tilde{V}$ and its capture cross-section $\sigma_c$. If we assume that the test-particle keeps a spherical shape on average and that each particle has volume $M / \tilde{\rho}$, where $\tilde{\rho}$ is the (proper) mass density of a dust grain, then the volume increases at a rate $\dot{\tilde{V}} = M \nu  / \tilde{\rho}$. Given $V = 4 \pi R^3/ 3$ we have
\be
\dot{R} = {\rho \over 4 \tilde{\rho}} \sqrt{(u \cdot U)^2 - 1}.
\ee
Since typically $\rho \ll \tilde{\rho}$ we expect the cross-section increase to be negligible.}
\par

\emph{Gravitational capture.}
In the case of gravitational capture one also must specifiy some inelastic mechanism under which scattering medium constituents (which start off unbound) become bound to the test-particle, i.e. the perturber. Contact with the event horizon or a hard physical surface are obvious candidates. In fact, accretion occurs at a much higher rate. As first pointed out by Hoyle and Lyttleton \cite{1939PCPS...35..405H}, incoming particles get focussed behind the perturber giving rise to a density wake. In this wake molecules are likely to collide, leading to loss of kinetic energy and for a portion of them to become gravitationally bound to the perturber.
This qualitatively explains why the gravitational capture cross-section should be much larger than the physical size of the perturber. 
\par
To get a quantitative estimate we compare the cross-sections for hard-sphere and gravitational scattering \eqref{eq:hs} and \eqref{eq:gcs}. We may assign an effective gravitational ``radius" to the perturber given by
\be
\label{eq:Rth}
R(\theta) =  {G m (1 + v^2)  \over  v^2 \sin^2(\theta / 2)}.
\ee
Note that we reinstated the perturber mass $m$, since $m$ is now dynamical.
\par
When arriving at the wake, medium constituents grazing closer to the perturber will have large tangential momentum which will be lost due to inelastic collision with the wake. The constituent will be captured if the remaining (radial) kinetic energy is smaller than the gravitational potential energy. Therefore, hard scattering angles, corresponding to smaller impact parameters, should lead to capture. 
\par
Hoyle and Lyttleton \cite{1939PCPS...35..405H} analyzed this problem in Newtonian mechanics where the relativistic factor $(1 + v^2)^2$ in the gravitational cross-section \eqref{eq:gcs} is absent. Their result is that capture occurs for $\theta \geq \pi/2$.\footnote{Therefore, in this model, accretion occurs for $\theta \in \left[{\pi \over 2}, \pi\right]$, while for  $\theta \in \left(\theta_\text{min}, {\pi \over 2}\right)$ matter gets gravitationally deflected, which leads to dynamical friction on the perturber. $\theta_\text{min}$ is fixed by $b_\text{max}$, the maximum impact parameter, ie. the length span of the medium.} Taking the same assumption for \eqref{eq:Rth} we find a ``capture radius"
\be
R_c = R\left({ \pi \over 2} \right) =  {2 G m (1 + v^2)  \over  v^2}.
\ee
Note that $R_c$ is much larger than the Schwarzschild radius for small velocities $v$, which is consistent with a large effective size of a gravitational perturber. The corresponding capture cross-section then reads
\be
\label{eq:capcg}
\sigma_c = \pi R_c^2 = {4 \pi G^2 m^2 (1 + v^2)^2  \over  v^4},
\ee
which, in the non-relativistic limit $v \to 0$, reduces to the Hoyle-Lyttleton expression \cite{1939PCPS...35..405H}.
\par
Plugging cross-section \eqref{eq:capcg} into eq. \eqref{eq:mdotacc} leads to an accretion rate given by
\be
\label{eq:HL}
\dot{m} = - 4 \pi G^2 \rho \, m^2 (u \cdot U)  { [2 (u \cdot U)^2 - 1]^2  \over   [(u \cdot U)^2 - 1]^{3 \over 2} }.
\ee
The corresponding covariant drag coefficient \eqref{eq:Bacc} reads
\be
\label{eq:HLB}
B = 4 \pi \rho \, G^2 m { [2 (u \cdot U)^2 - 1]^2  \over   [(u \cdot U)^2 - 1]^{3 \over 2} }.
\ee
Note the similarities with the dynamical friction coefficient \eqref{eq:Bdustg} where only the Coulomb logarithm $\Lambda$ is absent from the above and the mass $m$ is now evolving according to \eqref{eq:HL}.
\par

\section{Applications}
\label{sec:app}

In this section we apply some of our covariant formulas in specific curved backgrounds. We will mostly consider the Schwarzschild metric,
\be
\label{eq:BH}
ds^2 = - w(r) dt^2 +{1 \over w(r)} dr^2 + r^2 d \theta^2 +  r^2 \sin^2 \theta \, d\phi^2
\ee
with
\be
w(r) \equiv 1 - {2 G M \over r},
\ee
where $M$ is the mass of the black hole (BH).
\par

\subsection{Free-falling spring}
\label{sec:ffs}

Let us consider an elastic spring to which to one end we attach the test-particle and to the other end another massive object, the force applier. The system falls gravitationally in a generic curved background. We let test-particle and force applier be very close to each other so that the spring does not extend very much compared to the curvature scale of the background metric. We may then set a locally flat coordinate system \eqref{eq:normal} where we can make use of the Lorentz covariant expression for Hooke's law \eqref{eq:H}.
\par
Following the discussion at the start of section \ref{sec:lcv}, the covariant formulation of Hooke's law \eqref{eq:H}, is only expected to be physically accurate for a force applier in uniform motion in flat-space time. According to the equivalence principle, this generalizes, in curved space-time, to the free-falling condition for the force applier,
\be
\label{eq:DU}
{D U^\mu \over d \tilde{\tau}}= 0,
\ee
where $U^\mu$ is the 4-velocity of the force applier with proper time $\tilde{\tau}$. In practice, we may further assume the force applier to be a much heavier object than the test-particle, meaning that the force applier is unaffected by the elastic reaction of the spring and remains in free-fall. For example, the force applier can be a free-falling spaceship to which a spring is somewhere attached inside.
\par
If there was no spring, the test-particle would also have to be in free-fall, and $x^\mu$ would be the so-called \emph{deviation vector} whose evolution would be determined by the geodesic deviation equation \cite{Misner:1973prb}. Instead, the test-particle is forced out of its geodesic by the elastic force.  The geodesic deviation equation will need correction due to the spring.
\par
Let us find this correction explicitly. We start by noting that if $x^\mu = 0$, particle and applier will sit on top of each other and the force \eqref{eq:H} vanishes. The elastic force is thus like a tidal force in this regard (we will confirm this explicitly in a second). In this case both objects will follow the same geodesic, which implies
\be
\label{eq:uU}
u^\mu = U^\mu + O(x^\mu).
\ee
Thus, at leading order in $x^\mu$, the elastic 4-force reads
\be
\label{eq:FRH}
\F^{\mu \nu} u_\nu = k (x^\mu U^\nu - x^\nu U^\mu) U_\nu = - \bar{R}^\mu_{\; \, \alpha \beta \nu} U^\alpha x^\beta U^\nu
\ee
with
\be
\label{eq:RH}
\bar{R}_{\mu \alpha \beta \nu}  = - k (g_{\mu \beta} g_{\alpha \nu} - g_{\mu \alpha} g_{\beta \nu}),
\ee 
which one may recognize has the form of the Riemann tensor of Anti-de Sitter (AdS) space-time with
\be
\label{eq:AdSr}
\text{AdS radius} = {1 \over \sqrt{k}}.
\ee
\par
Plugging, \eqref{eq:FRH} into the equations of motion \eqref{eq:law} and letting the applier be in free-fall, eq. \eqref{eq:DU} we find, at leading order in $x^\mu$,
\be
\label{eq:gde}
{D^2 x^\mu \over d \tau^2} + (R^\mu_{\;\, \alpha \beta \nu} + \bar{R}^\mu_{\;\,\alpha \beta \nu})U^\alpha x^\beta U^\nu = 0,
\ee
where $R^\mu_{\;\, \alpha \beta \nu}$ is the Riemann curvature tensor of the space-time metric $g_{\mu \nu}$ \cite{Misner:1973prb} evaluated at the applier's worldline. When $k = 0$, the above reduces to the geodesic deviation equation.
\par
We see that for small displacements the elastic force can be interpreted as an AdS tidal force. This is not surprising because AdS geometry embodies many properties of the (Newtonian) harmonic potential. Free-falling particles in AdS follow harmonic motion \cite{Tho:2016fgz}, in the same way that for small amplitudes, $x^\mu \to 0$, the relativistic harmonic oscillator becomes non-relativistic and therefore harmonic \cite{1957AmJPh..25..535M,1965Natur.205..892H,PhysRevD.6.1474,1994AmJPh..62..531M}.
\par
Importantly, $g_{\mu \nu}$ in eq. \eqref{eq:RH} is an arbitrary metric. It is not the AdS metric with radius \eqref{eq:AdSr}. The metric $g_{\mu \nu}$ contributes with its own curvature $R^\mu_{\;\, \alpha \beta \nu}$ to the deviation equation \eqref{eq:gde}.

\par
For example, in a de Sitter universe we have
\be
\label{eq:dS}
R_{\mu \alpha \beta \nu}  = {\Lambda \over 3} (g_{\mu \beta} g_{\alpha \nu} - g_{\mu \alpha} g_{\beta \nu}),
\ee 
where $\Lambda$ is the cosmological constant. Plugging into \eqref{eq:gde} we see that for
\be
k \geq {\Lambda \over 3}
\ee
the test-particle and force applier will not be spread apart by the cosmological expansion.\footnote{Using concrete numbers taken from \cite{Planck:2018vyg} we have $\Lambda \sim 10^{-35} \, \text{s}^{-2}$ meaning, unsurprisingly, that an electrically bound electron-proton system, for which $k \sim 10^{33}  \, \text{s}^{-2}$, or the Earth-Sun system for which $k \sim 10^{-14}  \, \text{s}^{-2}$ would remain bound.}
\par
Let us now consider the spring free-falling radially onto a Schwarzschild BH, as described by the metric \eqref{eq:BH}. A radial geodesic will have 4-velocity given by \cite{Misner:1973prb}
\be{}
U^\mu(\tau) = \left({1 \over 1 - 2GM/r}, \sqrt{2GM \over r} , 0 , 0 \right)
\ee
where the radius $r=r(\tau)$ goes from $\infty$ to $0$ as $\tau$ increases (its precise dependence on $\tau$ does not concern us here). We now let
\be
x^\mu = (x^0, x_\parallel, x_\perp, 0)
\ee
and fix $x^0$ so that $x^\mu$ is orthogonal to $U^\mu$, i.e. so that in the free-falling observer's rest-frame $U^\mu = (1,\vec{0})$, $x^\mu$ is a purely spatial vector.\footnote{This choice is in fact arbitrary and preserved for any $\tau$ as $d(x^\mu U_\mu) / d\tau = 0$, which follows from contracting \eqref{eq:gde} with $U^\mu$, using eq. \eqref{eq:DU} with $d \tilde{\tau} \approx d \tau$ and antisymmetry on the first two indices of the Riemann tensor.}
\par
Note that $x_\parallel$ is the deviation along the radial direction while $x_\perp$ is the deviation orthogonal to the radial direction, which we have chosen, without loss of generality, to be along $\theta$. 
\par
We now make use of the Riemann tensor components of the Schwarzschild metric, which can be found in any textbook \cite{Misner:1973prb}, on eq. \eqref{eq:gde}. We find, by components,
\be
\label{eq:gdeBH}
{D^2 x_\parallel \over d \tau^2} = \left({2 G M \over r^3} - k \right) x_\parallel, \;\; {D^2 x_\perp \over d \tau^2} = - \left({G M \over r^3}+k \right) x_\perp.
\ee
First we note that there is no evidence of the event horizon here. Indeed, nothing special should happen as the spring crosses the event horizon, as it is a coordinate singularity. We also see that there is a `stretching' tidal force along the radial direction, while on the orthogonal direction there is a `compressing' tidal force. The latter will add to the restitution force of the spring while the former will compete with it. A comoving observer that sets up a local inertial frame, i.e. a passenger of the free-falling spaceship, will see the spring oscillate with frequencies\footnote{Note that this is not case in Schwarzschild coordinates as ${D^2 x^\mu \over d \tau^2} \neq \ddot{x}^\mu$. This is however indeed the case in the comoving free-falling frame, i.e. in a Fermi normal coordinate frame adapted to the radial geodesic \cite{1963JMP.....4..735M}. Applying the Fermi normal metric, eq. (79) of \cite{1963JMP.....4..735M}, and $U^\mu = (1, \vec{0})$, because a free-falling observer is at rest in this frame, into eq. \eqref{eq:gde} leads to \eqref{eq:gdeBH} with indeed ${D^2 x^\mu \over d \tau^2} = \ddot{x}^\mu$. The tidal forces remain invariant, which can be explained from their invariance under Lorentz boosts (see sec. 31.2 of \cite{Misner:1973prb}), and eqs. \eqref{eq:w} follow.}
\be
\label{eq:w}
\omega_\parallel = \sqrt{{2 GM \over r^3} - k }, \qquad \omega_\perp = \sqrt{{GM \over r^3} + k }.
\ee
We see that along the radial direction the spring will no longer oscillate after a critical radius
\be
r < r_c = \left({2 G M m \over k} \right)^{1 \over 3}.
\ee
where we reinstated the mass $m$. Once the spring goes below this radius it will be inexorably spread apart by the increasing gravitational tidal forces until it breaks. Taking typical values $m \sim 1 \text{ kg}$, $k \sim 1 \text{ N} / \text{cm}$, we find, for a stellar mass BH, that this happens at 
\be
r_c \sim 200 \, r_s
\ee
where $r_s  \sim 10 \text{ km}$, the Scwharzschild radius of a stellar mass BH.

\subsection{Black hole infall with drag}
\label{sec:bhid}

\begin{figure*}
\centering
\includegraphics[width=\textwidth]{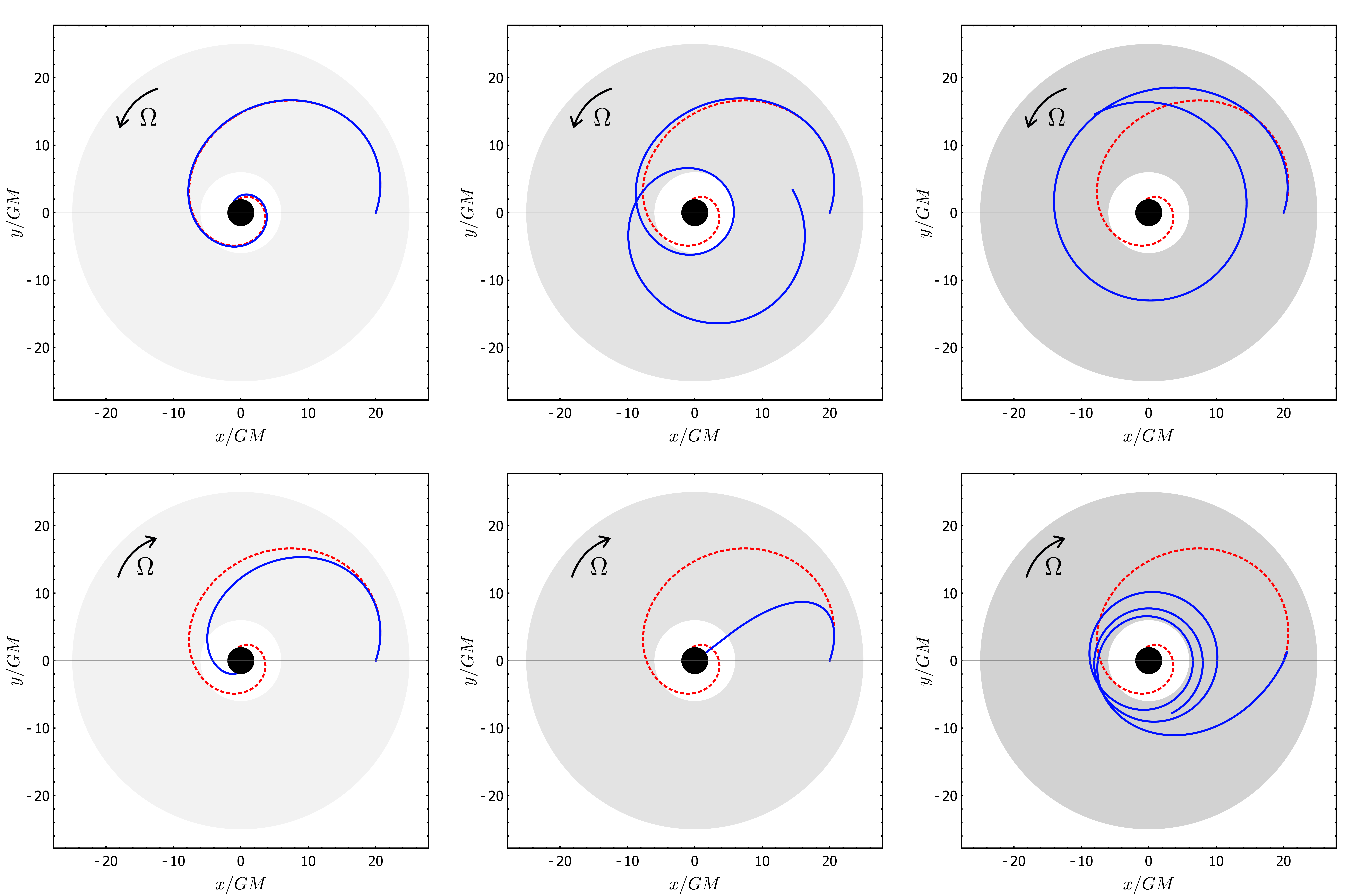}
\caption{Equatorial orbits in a Schwarzschild background with an accretion disk (in gray). In units where $G=c=M=1$ the initial data is $r(0) = 20$, $r'(0) = 0.06$, $\phi(0) = 0$, $\phi'(0) = 0.009$, $t(0)= 0$ and $t'(0)$ is fixed by the mass-shell condition $u^2(0) = -1$. The corresponding geodesic is plotted in dashed red. Trajectories obtained by numerically solving \eqref{eq:eombh} until $\tau = 500$ are plotted in thick blue. From left to right: $\rho_0 \sigma / m = 10^{-3},10^{-2}, 10^{-1} $. In the top row the disk is rotating counterclockwise (`$+$' sign in eq. \eqref{eq:acU}), while in the bottom row the disk is rotating clockwise (`$-$' sign in eq. \eqref{eq:acU}).  } \label{fig:orbits}
\end{figure*}

We again consider a Schwarzschild background \eqref{eq:BH} on which dust-like accretion orbits. We let an infalling test-particle be dragged by collisions with the dust constituents and make use of the hard-sphere dust drag coefficient \eqref{eq:Bdust} in eqs. \eqref{eq:2Fc} and \eqref{eq:eom}, leading to the following equations of motion,
\be
\label{eq:eombh}
\dot{u}^\mu + \Gamma^\mu_{\; \alpha \beta} u^\alpha u^\beta = {\rho \sigma \over m}\sqrt{(u \cdot U)^2 - 1} ( u^\mu (u \cdot U) + U^\mu).
\ee

For simplicity, we consider motion in the equatorial plane $\theta = \pi /2$, i.e. the test-particle is always immersed inside the accretion disk.
\par
In a dust model there are no inter-particle collisions, and therefore no shear stress along the disk.\footnote{This is a good approximation for thin disks \cite{frank}.} The disk constituents will then follow circular geodesics, with 4-velocity \cite{Misner:1973prb}
\be
\label{eq:acU}
U^\mu(r) = {1 \over \sqrt{1 - {3 G M \over r}}}\left(1,0,0, \pm \, \Omega(r) \right),
\ee
and Keplerian angular speed
\be
\Omega(r) = \sqrt{G M \over r^3}.
\ee
The closest possible circular orbit is at the photosphere where $r = 3 G M$. It is however the innermost \emph{stable} circular orbit (ISCO), $r = 6 G M$, that establishes the inner boundary of the accretion disk. We may thus consider the simplified density profile
\be
\rho(r) = \rho_0 \Theta(r - 6 G M) .
\ee
\par
An obvious point is that if the test-particle is co-moving with the disk, $u^\mu = U^\mu$, we see that the RHS of eq. \eqref{eq:eombh} vanishes and the test-particle will follow a geodesic. In particular, a circular geodesic with 4-velocity given by eq. \eqref{eq:acU}. A deviation of $u^\mu$ from $U^\mu$ leads to a net force on the test-particle, which eq. \eqref{eq:acU} tends to dynamically minimize. As we confirm numerically, at late times, any orbit (that does not end up in the singularity) becomes circularized, i.e. the test-particle joins the accretion flow. 
\par
In figure \ref{fig:orbits} we plot some trajectories (in blue) obtained from numerically solving eq. \eqref{eq:eombh}. The initial conditions were chosen as to lead to a geodesic that ends up in the singularity (plotted in dashed red). The parameter $\rho_0 \sigma / m$ takes the values $\rho_0 \sigma / m = 10^{-3},10^{-2}, 10^{-1}$ from left to right,\footnote{For comparison, taking data from \cite{Graham:2020gwr}, namely $\rho_0 \sim 10^{-10} \text{g cm}^{-3}$ and $M \sim 100 M_\odot$, and letting the test-particle be a spherical asteroid of radius $R$ and mass density $\sim 1 \text{ g cm}^{-3}$ \cite{2012P&SS...73...98C} we have $\rho_0 \sigma / m \sim 10^{-10}/ R$, which for an $R \sim 1$ cm asteroid amounts to $\rho_0 \sigma / m \sim 10^{-3}$.} while top and bottom rows have opposite rotations (`$\pm$' signs in eq. \eqref{eq:acU}) for the accretion disk.
\par
As expected, we observe stronger deviation from geodesic motion as $\rho_0 \sigma / m$ increases (left to right), and for retrograde motion of the disk with respect to the initial condition (bottom row). In the latter case, drag reduces the orbital velocity of the infalling particle leading to faster plunge for $\rho_0 \sigma / m = 10^{-3},10^{-2}$ (bottom left and center). However, for $\rho_0 \sigma / m = 10^{-1}$ (bottom right), the accretion disk immediately reverses the orbital velocity of the particle, leading to a retrograde stable orbit which circularizes at a radius $\sim 8 G M$.
\par
Similarly, in the top row, we observe that drag prevents the plunge of the particle for $\rho_0 \sigma / m = 10^{-2},10^{-1}$ (top center and right). In the top center case, the particle grazes the ISCO and has a highly eccentric orbit, which becomes circularized and co-moving with the disk after $\sim 18$ revolutions at a radius of $\sim 10 G M$. In the top right case, however, due to stronger wind, circularization of the orbit occurs sooner after $\sim 5$ revolutions and at the larger radius $\sim 14 G M$.

\section{Conclusion}
\label{sec:conclusion}

\begin{table*}
\begin{tabular}{ |c|c|c|c|c| }

\hline
$B(\gamma \equiv - u \cdot U)$ & Dust & Radiation & Gas (hot) & Gas  \\ \hline
Hard-sphere  &  $\rho \sigma \sqrt{\gamma^2 - 1}$ & $e \sigma \gamma$ &  ${4 \over 3} \rho \sigma \sqrt{8 / \pi \beta}$ &  Eq. \eqref{eq:bNhs}\\ \hline
Gravitational & $4 \pi \rho \, G^2 m^2 \Lambda [2 \gamma^2 - 1]^2 [\gamma^2 - 1]^{-{3 \over 2}} $ & $16 \pi e \, G^2 m^2 \Lambda \, \gamma$ &  $\frac{4}{3}  \rho \, G^2 m^2 \Lambda \,  \sqrt{2 \pi } \beta ^{3/2}$ & Eq. \eqref{eq:Bgasg} \\ \hline

\end{tabular}

\caption{Covariant drag coefficient $B$ by interaction (row) and medium (column). In the table, $\rho$ stands for the proper mass density and $e$ stands for the radiation energy density in the source (e.g. a star) rest frame. $\sigma$ stands for the sphere cross-section, $G$ the gravitational constant, $\Lambda$ the Coulomb logarithm which, depending on the effective size of the test-particle, may also depend on $\gamma$. Finally, $\beta \equiv M/ k_B T$ with $M$ the mass of a gas molecule, $k_B$ the Boltzmann constant and $T$ the gas temperature. In the case of radiation, the medium is massless, so $U^\mu$ is a null vector. The expressions for the gas were computed using the Maxwell-Boltzmann distribution. In the hot gas limit, i.e. when the thermal velocity largely exceeds the gas average speed, also known as Epstein regime \cite{PhysRev.23.710}, the drag coefficient becomes constant for either interaction. Several entries correspond to known forces. Namely, Petrich et al's dynamical friction \cite{Petrich} (gravitational, dust), Chandrasekhar's dynamical friction \cite{Chandra} (gravitational, gas), Poynting-Robertson drag \cite{1904RSPTA.202..525P,1937MNRAS..97..423R} (hard-sphere, radiation) and Epstein drag \cite{PhysRev.23.710} (hard-sphere, gas (hot)).}
\label{table:B}

\end{table*}

In this work we proposed a covariant framework for relativistic forces. Our goal was to find a general expression for the 4-force $f^\mu$ which satisfies the following requirements.
\begin{enumerate}
    \item $f^\mu$ is manifestly covariant.
    \item $f^\mu$ is such that $f \cdot u = 0$ as to preserve the mass-shell $u^2 = -1$.
    \item $f^\mu$ is fixed in terms of the force, via a Newtonian correspondence. 
\end{enumerate}
In sections \ref{sec:eom} and \ref{sec:covmap} we showed that
\be
\label{eq:final}
f^\mu = U^\mu (F \cdot u) - F^\mu (U \cdot u)
\ee
obeys all three requirements. Compliance with requirements 1 and 2 is evident. Requirement 3 was shown to hold via the equivalence principle. Concretely, in the local Lorentz frame where $U^\mu = (1, \vec{0})$, Newton's second law is recovered if $\vec{F}$ is identified with the force. As argued in section \ref{sec:covmap}, this uniquely fixes $f^\mu$.
\par
To make practical use of \eqref{eq:final}, the 4-vector $U^\mu$, which is unit and timelike, should be identified with the 4-velocity of some physical object. It follows, by the previous argument, that $\vec{F}$ is the force on the Lorentz rest-frame of that object. In most situations there are two possibilities for such an object. Either it is the test-particle itself, in which case $u^\mu = U^\mu$, or a second object which is responsible by the force on the test-particle, which we call the \emph{force applier}.\footnote{The force applier may also be seen as the test-particle itself, as in the case of self-forces (see appendix \ref{app:ald}).} 
\par
In section \ref{sec:lcv} we applied eq. \eqref{eq:final} to conservative forces. Concretely, we found covariant generalizations of: Coulomb's law, given by eq. \eqref{eq:C}, matching the Faraday tensor of a point-charge with 4-velocity $U^\mu$; Hooke's law, given by eq. \eqref{eq:H}, for which Gron's \cite{doi:10.1119/1.12623} analysis of how the spring constant $k$ would change in different inertial frames was reproduced (eqs. \eqref{eq:Gron} and \eqref{eq:Gron2}); and the constant force, which we considered to be of electrical origin (in order to identify the force applier as the infinite charged flat sheet that produces it). The corresponding Faraday tensor was obtained and is given by eq. \eqref{eq:cteC}.
\par
An important remark is that the obtained formulas are Lorentz covariant but \emph{not} general covariant. This is evident from the fact that they depend on quantities, such as the position $x^\mu$ of the test-particle, which, strictly, are 4-vectors only on Minkowski space-time. Such quantities involve two separate points, viz. the test-particle and force applier's positions, which have their own tangent spaces on a curved manifold. Fortunately, when both objects are sufficiently close to each other local flatness applies and Lorentz covariant expressions retain some use. This is shown explicitly in section \ref{sec:ffs} where the correction to the geodesic deviation equation were computed due to an elastic spring attached between neighbouring ``free-falling" wordlines. We considered the simple case of a radially free-falling spring, but it would be interesting to study other, more general, solutions of eq. \eqref{eq:gde}, e.g. as done in  \cite{bazanski1989geodesic,Kerner:2001cw}.
\par
One may wonder whether the Lorentz covariant expressions obtained for the conservative forces in section \ref{sec:lcv} may written in a general covariant way, i.e. for a generic curved background. Certainly, the relative position $x^\mu$ of the test-particle with respect to the force applier must be a replaced by a general covariant object. One natural possibility is to connect the test-particle and force applier via a geodesic, in which case the geodesic length would be the distance between both objects. There are however many geodesics that connect both worldlines. Which geodesic is the most appropriate?\footnote{It might be easier to generalize instead the \emph{potentials} (see appendix \ref{app:lagrangian}), since they are scalar invariants. }
\par
Another issue with at-a-distance forces is when the force applier is accelerated, e.g. as due to backreaction by its own effect on the test-particle. Every Newtonian force that was covariantly generalized in section \ref{sec:lcv} should be physically inaccurate in this case, i.e. if $\dot{U}^\mu \neq 0$ (this is surely the case for Coulomb's law). A retardation effect, as due to causality and the finiteness of the speed of light, should be accounted for. However, as past efforts indicate (see introduction, section \ref{sec:intro}), in particular the no-interaction theorem \cite{RevModPhys.35.350,currie1963interaction,doi:10.1063/1.1704121, leutwyler1965no}, a pure particle description of relativistic two-body systems is ill-fated. 
\par
Fortunately, these technical and fundamental hurdles essentially go away for most forces of astrophysical relevance, which are dissipative forces. Dissipative forces typically arise due to relative motion of a test-particle with respect to some medium which ``drags" its motion. Since this effect is local, i.e. as due to direct contact, there is no distance involved and, as shown in section \ref{sec:cf}, eq. \eqref{eq:final} yields expressions which are general covariant and valid for variable 4-velocity $U^\mu$ of the medium.\footnote{In fact, fundamentally, there is no such thing as a contact interaction. The  assumption at work is that the interaction between test-particle and medium constituents is of much shorter range as compared to the background curvature scale so that the equivalence principle applies. This is the case for virtually all aerodynamic-like drag forces, which are of electromagnetic origin, and dynamical friction on small enough compact objects, as in EMRIs.}
\par
Nonetheless, caution must be used in covariantly generalizing Newtonian drag forces, as they are velocity depend and may not hold relativistically. For this reason, before inserting them into eq. \eqref{eq:final}, in subsection \ref{ssec:scat} we re-derived a class of drag forces using relativistic kinematics. Concretely, master formula \eqref{eq:scatf} gives the drag on the test-particle's instantaneous rest frame due to collisions off medium constituents. We considered two astrophysically relevant interactions - hard-sphere and gravitational collisions - with corresponding differential scattering cross-sections, eqs. \eqref{eq:hs} and \eqref{eq:gcs}, in three different media: dust, radiation and gas. The corresponding covariant generalizations are summarized in table \ref{table:B}. 
\par
Many well-known cases are recovered. Namely, hard-sphere scattering on radiation leads to the Poynting-Robertson 4-force \cite{1937MNRAS..97..423R}. Gravitational scattering on dust and gas leads, respectively, to the covariantly generalized versions of Petrich et al's \cite{Petrich} and Chandrasekhar's \cite{Chandra} expressions. Hard-sphere scattering on gas leads to the covariant generalization of Epstein's result \cite{PhysRev.23.710}. Drag due to gravitational scattering on radiation and hard-sphere scattering on dust do not match to any well-known force. Moreover, note that none of the dust or gas drag coefficients is functionally identical to Poynting-Robertson drag (hard-sphere, radiation), which has been used to describe the latter \cite{Bini:2013pui,Bini:2013es,Bini:2016tqz}.  The use of the hard-sphere dust drag coefficient (hard-sphere, dust) was illustrated in section \ref{sec:bhid} for the case of an infalling test-particle into a Schwarzschild black hole with an accretion disk. 
\par
We note also that the gas drag coefficients were computed using the Maxwell-Boltzmann distribution, which only applies for a non-relativistic gas, i.e. with both non-relativistic temperature and average velocity. The former is typically the case for most astrophysical gases \cite{2011hea..book.....L}, while the latter may not be true depending on how fast a test-particle moves with respect to the gas. Even though this regime can be written covariantly (eq. \eqref{eq:nonrelreg}) it would be interesting to re-do the computation using a relativistic distribution.\footnote{One possibility is to use the Maxwell–Jüttner distribution (see appendix \ref{app:relgas}).}
\par
In subsection \ref{sec:vms} we considered variable-mass effects. We showed that these effects can be quantified in the same way as drag forces. For the variable-mass-rocket we found the following ``drag" coefficient
\be
\label{eq:Bvmr2}
B = - {d m \over d \tau}\, (u \cdot U)^{-1},
\ee
where $dm / d\tau < 0$ is the rocket mass depletion rate and $U^\mu$ is the 4-velocity of the propellant, which is related to the co-moving propellant velocity $v_e$ by eq. \eqref{eq:ve}. The relativistic rocket equation \cite{forward1995transparent} is recovered in flat space-time. Moreover, we find that, under the right conditions, the rocket will feel constant acceleration, eq. \eqref{eq:propera}. Therefore, the variable-mass-rocket can covariantly define hyperbolic motion. Rindler's  \cite{rindler1960hyperbolic}  requirement of planarity, that amounted to two conditions, is replaced by the concrete choice of direction of fuel ejection.
\par
Taking instead $dm / d\tau > 0$ allows to describe the force due to accretion. We considered a relativistic Hoyle-Lyttleton model \cite{1939PCPS...35..405H} and found the following accretion rate
\be
\label{eq:dotm}
{d m \over d \tau} = 4 \pi\rho \, G^2   m^2{  \gamma  [2 \gamma^2 - 1]^2  \over   [\gamma^2 - 1]^{3 \over 2} }, \;\text{ with }\; \gamma  \equiv - u \cdot U.
\ee
Plugging into \eqref{eq:Bvmr2} leads to the corresponding drag coefficient due to Hoyle-Lyttleton accretion \eqref{eq:HLB}. The result replicates the dust dynamical friction coefficient (bottom left in table \ref{table:B}) apart from the Coulomb logarithm $\Lambda$.
\par
We have not considered accretion of radiation, given that photons do not self-interact (at least classically) and there is no other inelastic mechanism that facilitates gravitational capture. Accretion may still occur, albeit at a much smaller rate, via direct contact with the physical surface of the compact object or the event horizon, in the case of a BH. This effect was recently quantified in the context of dark matter accretion/dynamical friction \cite{Traykova:2021dua,Vicente:2022ivh} onto a BH, where dark matter was modelled as a free scalar field. One natural way to introduce  inelasticity would be to add a self-interaction term for the scalar field $\phi$ to the Lagrangian (e.g. $\lambda \phi^4$). One expects that for $\lambda \gg 1$ the accretion rate to increase dramatically.\footnote{In this regime, dark matter should resemble a strongly collisional fluid which, in a scattering experiment, would lead to negligible escape at infinity. By energy-momentum conservation this in turn would imply considerable mass rate increase $dm / d\tau$ on the BH. }
\par
Bondi accretion \cite{1952MNRAS.112..195B}, i.e. accretion on a gas/fluid at non-zero temperature was also not considered. In most astrophysical situations, accretion flows are hydrodynamical \cite{1983bhwd.book.....S}, for which  a kinetic theory approach, as taken in this work, does not exactly apply. Relativistic hydrodynamics \cite{rezzolla2013relativistic} should be used instead (see e.g \cite{Richards:2021zbr} for a recent study).

\noindent{\bf{\em Acknowledgements.}}
We would like to thank José Natário for collaboration at the early stages of this project. We thank  Enrico Barausse for clarifying aspects of his work to us. We are grateful to Vitor Cardoso, José Natário, and Rodrigo Vicente for useful discussions. We also thank an anonymous referee for valuable suggestions. Finally, we thank Amit Sever and Alexander Zhiboedov for their encouragement and support.

\appendix

\section{Abraham-Lorentz-Dirac force}
\label{app:ald}

The Abraham-Lorentz force reads \cite{jackson_classical_1999},
\be
\label{eq:AL}
\vec{F}_N = { q^2  \over 6 \pi } {d^2 \vec{v} \over dt^2},
\ee
while the relativistic generalization derived by Dirac \cite{1938RSPSA.167..148D} reads
\be
\label{eq:D}
f^\mu = { q^2  \over 6 \pi } ( \ddot{u}^\mu - u^\mu \dot{u}^2 ).
\ee
Eq. \eqref{eq:AL} is only strictly valid in the comoving frame of the point charge. Thus, direct application of \eqref{eq:presc} gives
\be
F^\mu =  {q^2  \over 6 \pi} \ddot{u}^\mu,
\ee
and eq. \eqref{eq:2F} reads
\be
\label{eq:sf}
\F^{\mu \nu} = {q^2  \over 6 \pi } (u^\mu \ddot{u}^\nu - u^\nu \ddot{u}^\mu )
\ee
where we used $U^\mu = u^\mu$, which we can also interpret as the test-particle as being its own force applier. Dirac's expression \eqref{eq:D} then follows from $f^\mu = \F^{\mu \nu} u_\nu$ and the relation $u \cdot \ddot{u} = - \dot{u}^2$ which comes from differentiating the mass-shell \eqref{eq:mass-shell} twice.
\par
A general covariant expression is obtained by replacing $\dot{u}^\mu \to {D u^\mu \over d \tau}$ and $\ddot{u}^\mu \to {D^2 u^\mu \over d \tau^2}$. However, this is not the correct expression for the electromagnetic self-force in curved space \cite{1960AnPhy...9..220D,Hobbs1968RadiationDI,Barack:2018yvs,Poisson:2011nh}. In curved space-time the Abraham-Lorentz comoving force \eqref{eq:AL} is not accurate. This is due to the fact that, in a curved background, the wave equation Green's function has support inside the light cone \cite{1960AnPhy...9..220D}. This is because radiation may scatter back to the point charge giving rise to so-called ``tail" contributions to the self-force. There is also the addition of the Ricci tensor to the electromagnetic wave equation \cite{Hobbs1968RadiationDI} which, in the presence of matter, should also be accounted for.

\section{Lagrangian formulation}
\label{app:lagrangian}

We have described conservative forces (section \ref{sec:lcv}) by covariantly generalizing the force $\vec{F}_N$. However, conservative forces, can be written in terms of a potential $\phi$ as $\vec{F}_N = - \vec{\nabla} \phi$. According to eq. \eqref{eq:presc} this implies
\be
\label{eq:FPhi}
 F^\mu = - \partial^\mu \Phi
\ee
with covariant potential fixed by
\be
[\Phi = \phi]_{\vec{U} = 0}.
\ee
Moreover, if, as usual, the potential $\phi$ is independent of time, we must have
\be
\label{eq:ndV}
U^\mu \partial_\mu \Phi = 0.
\ee
In this case, contracting the equation of motion \eqref{eq:eom} with $U^\mu$ leads to conservation of the \emph{mechanical energy},
\be
\label{eq:EM}
E_M = - u \cdot U + \Phi = \text{constant},
\ee
which in the force applier's rest-frame, where $U^\mu = (1,\vec{0})$, amounts to $E_M = 1/\sqrt{1 - v^2} + \phi$, where $v$ is the test-particle's velocity.
\par
The examples considered in section \ref{sec:lcv} have the following potentials,
\begin{align}
\Phi_\text{Coulomb} &= {q \over \sqrt{x^2 + (U \cdot x)^2}  },  \\
 \Phi_\text{Hooke} &= {k \over 2} \left[x^2 + (U \cdot x)^2 \right], \\
 \Phi_\text{Constant} &= - {\sigma \over 2} \left[n \cdot x + (U \cdot n) (U \cdot x) \right].
\end{align}
All the above examples satisfy \eqref{eq:ndV}. Also note that applying \eqref{eq:FPhi} to the above leads to the corresponding $F^\mu$ in section \ref{sec:lcv} with some extra term along $U^\mu$ which however do not contribute to $\F^{\mu \nu}$ in eq. \eqref{eq:2F}.

The equations of motion \eqref{eq:eom} involving a conservative force \eqref{eq:FPhi} follow, equivalently, from the Euler-Lagrange equations associated with the action
\be
\label{eq:S}
S = \int d\tau \left[ {m \dot{x}^2 \over 2} + ( \dot{x} \cdot U) \Phi  \right].
\ee
According to Noether's theorem \cite{rund1966hamilton,mann1974classical}, if $\tau \to \tau + \delta \tau$ and $x^\mu(\tau) \to x^\mu(\tau) + \delta x^\mu$ is a symmetry of the action, then the combination
\be
\label{eq:Noether}
L \, \delta \tau +  \left( {\partial L \over \partial \dot{x}^\mu } \right) \delta x^\mu = \text{constant}
\ee
is a constant of motion. It is clear that translations in $\tau$ are a symmetry of the action \eqref{eq:S}. This corresponds to $\delta \tau = \text{const}$ and $\delta x^\mu = - \delta \tau \, \dot{x}^\mu$, which plugging into the above yields the (super-)Hamiltonian,
\be
H \equiv \left( {\partial L \over \partial \dot{x}^\mu } \right) \dot{x}^\mu - L ={g_{\mu \nu} \dot{x}^\mu  \dot{x}^\nu \over 2} = \text{constant}.
\ee
Given the initial condition $H(0) = 1/2$ we see that the mass-shell \eqref{eq:mass-shell} is preserved.
\par
Conservation of the mechanical energy \eqref{eq:EM} comes up as a further symmetry of the Lagrangian. Note that if eq. \eqref{eq:ndV} is verified then the action \eqref{eq:S} is invariant under spacetime shifts along $U^\mu$, i.e. time-independence in the applier's rest frame. Plugging $\delta x^\mu = U^\mu$ and $\delta \tau = 0$ into eq. \eqref{eq:Noether} yields
\be
U^\mu \left( {\partial L \over \partial \dot{x}^\mu } \right) = - E_M = \text{constant},
\ee
 with $E_M$ given by eq. \eqref{eq:EM}.

\onecolumngrid

\section{Relativistic hard-sphere gas drag}
\label{app:relgas}

The Maxwell-Jüttner distribution \cite{1911AnP...339..856J,1963JMP.....4.1163I,osti_4752122} reads 
\be
\label{eq:MJ}
W(\vec{k}) = {n \beta \over 4 \pi K_2(\beta)} \, e^{\beta k \cdot U}
\ee
with $n$ the proper particle number density, $U^\mu$ the average 4-velocity of the gas, $K_\nu(\beta)$ the modified Bessel function of the second kind.\footnote{We use the ``rapidity" integral representation,
\be
\label{eq:bessel}
K_\nu(\beta) = {\sqrt{\pi} \over  \Gamma(\nu + {1 \over 2}) } \left({\beta \over 2}\right)^\nu  \! \int\limits_0^\infty e^{- \beta \cosh \chi} \sinh^{2 \nu}\!\chi \, d\chi.
\ee } 
For $\beta \gg 1$, as with most astrophysical gases, we have $K_2(\beta) \to \sqrt{\pi \over 2 \beta} e^{- \beta}$ and the integral \eqref{eq:scatfg} will be dominated by the saddle point of the exponent. Expanding $k \cdot U$ to quadratic order around its saddle point turns $W(\vec{k})$ into a gaussian,
\be
\label{eq:MJB}
W(\vec{k}) \approx n \left({ \beta \over 2 \pi } \right)^{3 \over 2}\exp\left[{- {\beta \over 2 } M_{i j} (k^i - U^i) (k^j - U^j) }\right]\!,
\ee
but not quite a Maxwell-Boltzmann distribution since 
\be
M^{i j} = \delta^{ i j} - {U^i U^j \over (U^0)^2} 
\ee
is not proportional to the identity, a feature which introduces anisotropy into the distribution of momenta. If the gas also moves non-relativistically on average, i.e. $|\vec{U}| \ll 1$ and $U^0 \to 1$, then $M^{i j} \to \delta^{i j}$ and \eqref{eq:MJB} turns into the Maxwell-Boltzmann distribution. Note, however, that contrarily to the Maxwell-Boltzmann distribution, eq. \eqref{eq:MJB} is Lorentz invariant (under the saddle-point approximation). In the zero temperature limit $\beta \to \infty$, distribution \eqref{eq:MJB} will go to the Lorentz invariant delta function
\be
{W(\vec{k}) \over n} \to  {\delta^3(\vec{k} - \vec{U}) \over \sqrt{\det M}} = k^0 \delta^3(\vec{k} - \vec{U}),
\ee
which leads eq. \eqref{eq:scatfg} into the dust expression eq. \eqref{eq:scatf}.
\par 
\par
Let us now compute the force due to hard-sphere collisions on a Maxwell-Jüttner gas at non-relativistic temperatures ($\beta \to \infty$). For calculational purposes we find it simpler to apply the saddle-point approximation $\beta \to \infty$ at a later stage of the computation. Plugging distribution \eqref{eq:MJ} into eq. \eqref{eq:scatfg} reads
\be
\vec{F}_N = {\rho\sigma \beta \over 4 \pi K_2(\beta)} \int {|\vec{k}| \vec{k} \over k^0}  e^{\beta k \cdot U} d^3 \vec{k}.
\ee
To compute this integral we use spherical coordinates for $\vec{k}$ with $\vec{U}$ aligned along the $z$ direction.  The integral over the azimuthal angle averages to zero the components of $\vec{k}$ orthogonal to $\vec{U}$. We then get a force along $\vec{U}$,
\be
\label{eq:rte}
\vec{F}_N = {\rho \sigma \beta \over 2 K_2(\beta) |\vec{U}|} I(\beta, \vec{U}) \, \vec{U},
\ee
with
\be
I(\beta, \vec{U}) = \int_0^\infty  e^{- \beta k^0 U^0}  |\vec{k}|^4 {d |\vec{k}| \over k^0} \int_{-1}^1   e^{\beta |\vec{k}| |\vec{U}| x}  x \, d x,
\ee
where $x$ is the cosine of the polar angle, the angle between $\vec{k}$ and $\vec{U}$. The integral over $x$ is trivial and leads to
\be
I(\beta, \vec{U}) = {1 \over \beta^2 |\vec{U}|^2} \int_0^\infty  \bigg[e^{- \beta k^0 U^0 + \beta |\vec{k}| |\vec{U}|}(\beta |\vec{k}| |\vec{U}|-1)  + e^{- \beta k^0 U^0 - \beta |\vec{k}| |\vec{U}|}(\beta |\vec{k}| |\vec{U}|+1)\bigg] |\vec{k}|^2 {d |\vec{k}| \over k^0}.
\ee
Parametrizing over the rapidities,
\begin{align}
\label{eq:UKetachi}
U^0 &= \cosh \eta, \qquad |\vec{U}| = \sinh \eta, \\
k^0 &= \cosh \chi, \qquad \,|\vec{k}| = \sinh \chi,
\end{align}
allows to rewrite the integral as
\be
\label{eq:Ichi}
I(\beta, \eta) = {1 \over \beta^2 \sinh^2 \eta} \int_0^\infty  \bigg[e^{- \beta \cosh(\chi - \eta)}(\beta \sinh \eta \sinh \chi -1) 
+ e^{- \beta \cosh(\chi + \eta)}(\beta \sinh \eta \sinh \chi +1)\bigg] \sinh^2 \!\chi \,d \chi.
\ee
We now shift integration variable, $\chi \to \chi + \eta$ and $\chi \to \chi - \eta$ in the first and second pieces, respectively, to get
\begin{align}
\label{eq:Ieta}
I(\beta, \eta) &= {2 \over \beta^2 \sinh^2 \eta} \bigg[ - \int_0^\eta d \chi \, e^{- \beta \cosh \chi} \sinh^2(\chi - \eta) \big( \beta \sinh \eta  \sinh(\chi - \eta) + 1\big) \notag \\ 
& \qquad + \, \sinh 2\eta \int_0^\infty \! d \chi \, e^{- \beta \cosh \chi} \sinh \chi \big(\beta \cosh^2 \eta \sinh^2 \chi - 2\cosh \chi + 3 \beta \sinh^2 \eta \cosh^2 \chi \big)
\bigg].
\end{align}
The shift of integration variables from \eqref{eq:Ichi} to \eqref{eq:Ieta} essentially amounts to a boost back to the rest frame of the gas, where it should follow a Maxwell-Boltzmann distribution, since the gas is assumed to be non-relativistic. Indeed, in the limit $\beta \to \infty$, both integrals will be dominated by the $\chi \to 0$ region due to the decaying exponential. In this limit we have
\be
\label{eq:chinonrel}
 \sinh \chi \to \chi, \qquad \cosh \chi \to 1, \qquad  e^{- \beta \cosh \chi} \to e^{- \beta(1 + {\chi^2 \over 2})}.
\ee
Making use of \eqref{eq:UKetachi} we then get
\begin{align}
\label{eq:IU}
I(\beta, \vec{U})  \approx {2 e^{- \beta} \over \beta^2 |\vec{U}|^2} \bigg[ - \int_0^\eta d \chi \, e^{- {\beta \chi^2 \over 2} } &\left(U^0 \chi - |\vec{U}|\right)^2 \left(1 - \beta |\vec{U}|^2  + \beta U^0 |\vec{U}| \chi \right) \notag \\ 
& \qquad +\, \, U^0 |\vec{U}| \int_0^\infty \! d \chi \, e^{- {\beta \chi^2 \over 2} } \chi \left(\beta (U^0)^2 \chi^2 - 2 + 3 \beta |\vec{U}|^2  \right)
\bigg].
\end{align}
Note that we cannot \emph{a priori} discard the higher powers in $\chi$ in the above because we don't know how $\chi$ compares to $|\vec{U}|$ or $U^0$. We wish to be completely agnostic about the magnitude of $|\vec{U}|$. Eq. \eqref{eq:IU} integrates to
\begin{align}
\label{eq:IUf}
I(\beta, \vec{U})  = {2 e^{- \beta} \over \beta^3 |\vec{U}|^2} &\bigg[  U^0  e^{- {\beta \eta^2 \over 2}} \left(3 \beta |\vec{U}|^3+ \eta U^0 - 3 \beta \eta U^0 |\vec{U}|^2  + \beta \eta^2 (U^0)^2 |\vec{U}| \right)\notag \\ 
&  \qquad + \,\sqrt{ \pi \over 2 \beta} \, \erf\bigg({\sqrt{\beta \over 2} \eta}\bigg) \left(\beta^2 |\vec{U}|^4  + 2 \beta  |\vec{U}|^2 - 1 \right)
\bigg].
\end{align}
Plugging back into \eqref{eq:rte} we get $\vec{f} = b \vec{U}$ with drag coefficient given by
\begin{align}
\label{eq:rgd}
b =  {\rho \sigma \over \beta^2 |\vec{U}|^3} \bigg[   \sqrt{2 \beta \over \pi} U^0  e^{- {\beta \eta^2 \over 2}} &\left(3 \beta |\vec{U}|^3+ \eta U^0 - 3 \beta \eta U^0 |\vec{U}|^2  + \beta \eta^2 (U^0)^2 |\vec{U}|  \right)\notag \\ 
&  \qquad + \,  \erf\bigg({\sqrt{\beta \over 2} \eta}\bigg)  \left(\beta^2 |\vec{U}|^4  + 2 \beta  |\vec{U}|^2 - 1 \right)
\bigg] .
\end{align}
The covariant drag coefficient $B$ then follows from
\be
|\vec{U}| = \sqrt{(U^0)^2 - 1}, \qquad \eta = \mathrm{arccosh} \, U^0, \qquad 
\ee
and replacing $U^0 \to - u \cdot U$ in the above.
\par
The computation with arbitrary $\beta$ is difficult, but in the limit where the gas is moving macroscopically slow, we were able to do it. As with Epstein's original approach \cite{PhysRev.23.710}, we directly expand at leading order in $\vec{U}$ the distribution function \eqref{eq:MJ}. The integral \eqref{eq:rte} can then be expressed in terms of a Bessel function and is given by
\be
b_\text{Epstein rel. gas} = {4 \over 3} \sqrt{8 \over \pi \beta} \rho \sigma {K_{5/2}(\beta) \over K_2(\beta)}.
\ee
For $\beta \to \infty$ the ratio of the Bessel functions goes to $1$ and we recover Epstein's result \eqref{eq:BEpstein}.

\twocolumngrid

\bibliography{References}

\end{document}